\documentclass[pra,twocolumn,showpacs,superscriptaddress]{revtex4}

\usepackage{graphicx,color,amsmath,amsfonts, subfigure}
\usepackage[all] {xy}
\graphicspath{{Diagrams/}}

\newcommand{\beq}{\begin{equation}}
\newcommand{\eeq}{\end{equation}}
\newcommand{\bqa}{\begin{eqnarray}}
\newcommand{\eqa}{\end{eqnarray}}

\newcommand{\erf}[1]{Eq.~(\ref{#1})}
\newcommand{\erfs}[2]{{ Eqs.~(\ref{#1})--(\ref{#2})}}

\newcommand{\crf}[1]{Ref.~\cite{#1}} 
\newcommand{\srf}[1]{Sec.~\ref{#1}}
\newcommand{\ea}{{\it et al.}}

\newcommand{\dg}{^\dagger}

\newcommand{\cu}[1]{\left\{ {#1} \right\}}
\newcommand{\ro}[1]{\left( {#1} \right)}

\newcommand{\xfrac}[2]{{#1}/{#2}}

\newcommand{\ket}[1]{\ensuremath{| #1 \rangle}}
\newcommand{\bra}[1]{\ensuremath{\langle #1 |}}

\definecolor{nblue}{rgb}{0.2,0.2,0.7}
\definecolor{ngreen}{rgb}{0.2,0.6,0.2}
\definecolor{nred}{rgb}{0.7,0.2,0.2}
\definecolor{nblack}{rgb}{0,0,0}

\begin{document}

\title{Tracking an open quantum system using a finite state machine: stability analysis}
\author{R. I. Karasik } 
\author{H. M. Wiseman}\email{H.Wiseman@Griffith.edu.au}
\affiliation{Centre for Quantum Computation and Communication Technology (Australian Research Council), \\ Centre for Quantum Dynamics, Griffith University, Brisbane, Queensland 4111, Australia}

\begin{abstract} 
A finite-dimensional Markovian open quantum system will undergo  quantum jumps between pure states, if we can monitor the bath to which it is coupled with sufficient precision.
In general these jumps, plus the between-jump evolution, create a trajectory which passes through infinitely many different pure states, 
even for ergodic systems. However, as shown recently by us [Phys. Rev. Lett. \textbf{106}, 020406 (2011)], it is possible to construct {\em adaptive} monitorings  
which restrict the system to jumping between a finite number of states. That is, it is possible to track the system using a {\em finite state machine} as the apparatus. In this paper we consider the question of the stability of these monitoring schemes. 
Restricting to cyclic jumps for a qubit, we give a strong analytical argument that these schemes are always stable, and supporting analytical and numerical evidence for the example of resonance fluorescence. This example also enables us to explore a range of 
behaviors in the evolution of individual trajectories, for several different monitoring schemes. 
\end{abstract}
 
\maketitle

\section{Introduction}

An open system is one which continuously exchanges information with its environment \cite{Car93b,BrePet02,WisMil10}. For 
a Markovian (memoryless) system, this amounts to a loss of information from the system into the environment. 
By {\em monitoring} the environment, it is possible to regain this lost information and hence to {\em track} the system. 
If the monitoring process is perfect, then one can expect to perfectly track the system; that is, to know as much as 
is possible to know about the system, as would be the case it were a closed system prepared at will.

These very general considerations apply equally to classical and quantum systems.
However, there are some very significant differences between the two cases. In the classical case, there is only one 
(best) way to monitor the environment. Also, if the classical system  
has only finitely many possible states --- this is known as a finite state machine \cite{Gil62} --- then obviously it is possible 
to keep track of its state using only a finite classical memory --- another finite state machine of the same dimension. 
In the quantum case, by contrast, there are infinitely many inequivalent ways of monitoring the environment that enable 
the experimenter to deduce what pure state the system is in \cite{Car93b,Wis96a,WisMil10}. This is because of the 
entanglement between the system and its environment. But in almost all cases, 
an infinite classical memory is required to store that pure state, even for a finite-dimensional quantum system. 

This last point can perhaps only be understood by introducing a little formalism. 
We consider finite-dimensional systems that undergo Markovian 
open quantum system dynamics, described by  a Lindblad-form master equation \cite{BrePet02,WisMil10}: 
\begin{equation}
\label{me1}
\dot{\rho} = {\cal L}\rho \equiv -i[  \hat H_{\rm eff}\rho   -  \rho\hat H_{\rm eff}\dg] 
+ \sum_{l=1}^L \hat c_l \rho \hat c\dg_l,
\end{equation}
where $ \hat H_{\rm eff} \equiv \hat H - i\sum_l  \hat c\dg_l \hat c_l/2$.  
Here $\hat H$ is Hermitian (it is the Hamiltonian) 
but the  jump operators $\{\hat c_l:l\}$ are completely arbitrary.
This decomposition defines the evolution of the system conditioned on a perfect 
monitoring of the bath quanta. If a quantum in channel $l$ is observed at time $t_n$, the system state jumps from   
the pre-jump state $\ket{\psi(t_n^-)}$ to the post-jump state $\ket{\psi(t_n)}\propto \hat c_l\ket{\psi(t_n^-)}$.
Then until the next jump its subsequent 
evolution would be generated by the effective (non-Hermitian) Hamiltonian $\hat H_{\rm eff}$ appearing in \erf{me1}. 

In general, the post-jump state will depend on the pre-jump state $\ket{\psi(t_n^-)}$, 
and it will not remain stationary until the next jump, unless it happens to be an eigenstate of $\hat H_{\rm eff}$. 
It is thus not at all obvious whether for a general open quantum system it would be possible to keep track of its pure state, even in principle,  with a finite classical memory. On the face of it, it would seem necessary to store the nature and exact times of each jump --- requiring a sequence of real numbers $\{t_n:n\}$ each of which would require, in principle, an infinite memory to store. Alternately one could store the conditioned quantum state   $\ket{\psi(t)}$ itself, but this (a finite vector of complex numbers)   would also require an infinite memory.

In Ref.~\cite{KarWis11} we showed that, for an arbitrary $D$-dimensional quantum 
system obeying a Markovian master equation with a unique stationary (mixed) state, one can expect there to exist a monitoring 
such that it is possible to track the system's  
conditional (pure) state with a $K$-state machine as apparatus, with some $ K \geq (D-1)^2+1$. For a qubit we proved that this is indeed always the case: a two-state apparatus can be found that ensures 
the qubit jumps between just two states. This apparatus must implement an {\em adaptive} monitoring, 
choosing how to measure the environment depending on its own internal state. This result shows that 
the infinite amount of information required to track a quantum system under a generic monitoring scheme 
arises from the poor choice of scheme and is not intrinsic to the coupling between the quantum system 
and its environment.

The goal of this paper is to investigate the stability of such finite-state monitoring and provide details on how such schemes can be identified and constructed. The first goal is necessary to establish that the schemes introduced in Ref.~\cite{KarWis11} are not just mathematical constructions, but are physically realizable. For this purpose we restrict to qubit evolution, and cyclic-jump schemes. 
All the schemes we study are stable in a mean-square sense, and we give a strong analytical arguments plus supporting numerical evidence
(for two- and three-state machines) that such schemes are always stable. 

The paper is organized as follows. In section \ref{sec:prefEnsemble} we review the background for this paper and summarize relevant results from Ref.~\cite{KarWis11}. In section \ref{sec:jumpDynamics} we develop our main results for stability. We discuss these results in the context of specific example of resonance fluorescence in section \ref{sec:Resfluor}. 
We conclude with a summary and a statement of open problems in this area.

\section{The preferred ensemble} 
\label{sec:prefEnsemble}
The idea of tracking the system with finite classical memory is closely related to the concept of a preferred ensemble. To explain the  latter notion, we consider an evolution generated by a Markovian master equation with unique steady state defined by ${\cal L}\rho_{\rm ss}=0$ and we assume that $\rho_{\rm ss}$ is a mixed state. 
This mixed state can be decomposed in terms of  an ensemble of  pure states $\ket{v_k^e}$ with some positive weights $\wp_k$ such that 
\beq
\rho_{\rm ss}= \sum_{k=1}^K  \wp_k \ket{v_k^e}\bra{v_k^e}.
\eeq
 Note that there are infinitely many such decompositions as the states $\ket{v_k^e}$ need not be orthogonal.  A valid   interpretation for any such decomposition of a mixed state  is  that in the long time limit a measurement performed on the environment will collapse 
the system into  one of the possible states $\ket{v_k^e}$ and this will happen with probability $\wp_k$. Such a measurement will, in general, require simultaneously measuring parts of environment that interacted with the system at {\em different} times in the past. 

A different question arises if we require 
the measurements on the environment to be continuous in time, so that the system state is continually being collapsed, and so (in the long time limit) will be in a stochastically evolving pure state. Because the system evolution is Markovian, this can be done while leaving unchanged the average evolution of the system  (\ref{me1}). Now the natural interpretation for the ensemble $\cu{\wp_k,\ket{v_k^e}}$ is that 
that the system will only ever be in one of the states $\ket{v_k^e}$, and will spend a proportion of time in that state equal to $\wp_k$. In this case, this interpretation is {\em not} valid for most ensembles. 
Decompositions  $\{\wp_k , \ket{v_k^e}\}$  that can be realized through an experiment via continuous measurement are called {\em physically realizable} (PR).  The fact that some ensembles are not PR  is known as {\em the preferred ensemble fact} \cite{WisVac01}. We note here that  if we know of a PR ensemble with $K$ finite, then if at some time $t$  the system is a pure state from the PR ensemble, $\ket{v_k^e}$, and is subject to  the continuous monitoring that realizes  this  PR ensemble, then  in its subsequent conditional evolution the system will only occupy states from the PR ensemble and we can therefore  track such evolution with  a  classical register  
with only $K$ states. Such a classical device is known as a finite state machine. 

From Ref.~\cite{WisVac01} we know that the ensemble $\{ \wp_k,\, \ket{v_k^e} \}$ is PR if and only if (iff) there exists rates $\kappa_{jk}>0$ such that 
\beq
\label{jumpCond} 
  \forall k,  {\cal L}\ket{v_k^e}\bra{v_k^e} = \sum_{k=1}^K \kappa_{jk} 
 \left(\ket{v_j^e} \bra{v_j^e}-\ket{v_k^e} \bra{v_k^e} \right),
\eeq
where the system jumps between $K$ different states. 
Typically, most ensembles $\{ \wp_k,\, \ket{v_k^e} \}$  representing $\rho_{\rm ss}$ are not PR, including the $K=D$ ensemble composed from the diagonal basis for $\rho_{\rm ss}$ \cite{WisMil10}.

For a qubit,  the  conditions in Eq.~(\ref{jumpCond}) can be simplified further. Using the Bloch representation,  \erf{me1} becomes
\beq
\dot{\vec{r}}=A\vec{r}+\vec{b},
\label{bve}
\eeq 
where $A$ is  a $3\times3$-matrix that dictates the evolution of the state and $\vec{b}$, a 3-vector, determines  the steady state of the system, $\vec{r}_{\rm ss}=-A^{-1}\vec{b}$.  This equation has a unique steady state  iff (if and only if)  the real parts of all eigenvalues of $A$ are negative. We can track this system with a $K$-state memory iff there exists  an ensemble $\{\wp_k,\,\vec{r}_k\}$ and rates $\kappa_{jk}\geq 0$ such that 
\begin{align}
\vec{r}_k\cdot\vec{r}_k=&1\,\,\forall k \label{bvNormCond}\\
A\vec{r}_j+\vec{b}=&\sum_{k=1}^K \kappa_{jk}(\vec{r}_k-\vec{r}_j)\,\,\forall j.
\label{bvJumpCond}
\end{align}
These equations are better than conditions from Eq.~(\ref{jumpCond}) for numerical search for PR ensembles because they generically reduce the number of different equations, the number of different variables as well as the maximal degree of this system of equations. Thus search for PR ensembles reduces to finding a real solution to a system of quadratic polynomials with real coefficients. Unfortunately, this is still a hard problem, which in general is known to be NP-complete \cite{BCSS}.  Equations~(\ref{bvNormCond})--(\ref{bvJumpCond})  are used in  \srf{3stateJump}  to analyze PR ensemble for $K=3$ for a qubit.

For the moment, we concentrate on a $K=2$ PR ensemble. In Ref.~\cite{KarWis11}, we showed that such  ensemble always exists for a qubit. This can be easily deduced from  \erfs{bvNormCond}{bvJumpCond},  which for $K=2$ imply that $A(\vec{r}_1-\vec{r}_2)=(\kappa_{12}+\kappa_{21})(\vec{r}_2-\vec{r}_1)$. This is an eigenvalue equation and we can conclude that 
\begin{align}
\vec{r}_1=&\vec{r}_{\rm ss}+\eta_1\hat{u} \label{r1e}\\
\vec{r}_2=&\vec{r}_{\rm  ss}-\eta_2\hat{u} \label{r2e},
\end{align}
where $\hat{u}$ is the normalized eigenvector of matrix $A$ and parameters $\eta_1$ and $\eta_2$ relate to probabilities $\wp_1$ and $\wp_2$ for occupying state $\vec{r}_1$ and $\vec{r}_2$, respectively, and $\wp_j=\eta_j/(\eta_1+\eta_2)$. Details expressions for $\eta_1$ and $\eta_2$  can be found in \crf{KarWis11},  but are not relevant to this paper. Here we just need the general structure of the solutions. 
Because the Bloch vectors, $\vec{r}_1$ and $\vec{r}_2$, must be real, only real eigenvectors $\hat{u}$ of $A$ can contribute to the solution.  As  a $3\times3$-matrix, $A$ has three eigenvalues and, by fundamental theorem of algebra, at least one eigenvalue (and consequently one eigenvector) is real. Therefore, a qubit always has a preferred ensemble comprising just two states and there can be up to three different solutions.

Different PR ensembles (including three different PR ensembles to track a qubit with one bit \cite{KarWis11}) arise from the freedom that experimentalists have (in principle) to monitor the system's environment in different ways. This freedom exists because of the invariance properties of the master equation, Eq.~(\ref{me1}), which is invariant with respect to transformations
\begin{align}
 \cu{\hat c_l} &\to \cu{\hat c'_m = \sum_{l=1}^{ L} S_{ml}\hat c_l + \mu_m  }  \label{trans1}\\
\hat H  &\to \hat H'= \hat H - \frac{i}{2}\sum_{m=1}^{M}  (\mu^*_m  \hat c'_m - \mu_m \hat c'_m{}\dg) \label{trans2}
\end{align}
where $\vec\mu=(\mu_1, \ldots,\mu_M)$ is an arbitrary complex vector and 
${\bf S}$ is an arbitrary semi-unitary  matrix --- $\sum_{m=1}^{M} S_{l'm}^* S_{ml} = \delta_{l',l}$.  	
Realizations of this master equation with $\cu{\hat c'_m}$ as the jump operators and $\hat H'_{\rm eff} = \hat H' - i \sum_{m=1}^M \hat c'_m{} \dg \hat c'_m /2$ as the effective non-Hermitian Hamiltonian generate the same average evolution as the original master equation, but clearly give rise to different stochastic evolution. 
To obtain the most general pure-state unravelling of the master equation, we require $\vec\mu$ and ${\bf S}$ to depend upon the previous record of jumps. That is, we require an adaptive monitoring. Of course when we use this to achieve jumping between a finite number of states, the classical $K$-state memory that stores which state the system is currently in carries all the information necessary for determining $\vec\mu$ and ${\bf S}$. That is, the adaptive unravelling is specified by $k$ different values for $\vec{\mu}$ and ${\bf S}$. 

The physical meaning of these parameters is most easily explained in a quantum optics context.   The matrix 
 ${\bf S}$ describes a linear interferometer  taking the field outputs from the system as inputs and interferes them prior to detection. 
 This  is to be understood in the most general sense, including frequency shifters if the system has outputs in different frequency bands. 
The vector $\vec\mu$ describes adding (weak) local oscillators to the output fields from the interferometer prior to detection by photon counting. There have been for many years theoretical proposals for adaptively controlling the local oscillator amplitude \cite{Dol73} or 
phase \cite{Wis95c,Wis96a}, and more recently a number of experiments have been performed  \cite{Arm02,JM07,Whe10}, one of which 
(\crf{JM07}) used a weak local oscillator, with amplitude comparable to that of the system.

One characteristic that sets apart different solutions is the Shannon entropy. Under continuous monitoring, the system will occupy states $\ket{v_k^e}$ with probabilities $\wp_k$. The Shannon entropy for an ensemble $\{\wp_k, \ket{v_k^e}\bra{v_k^e}\}$ that represents $\rho_{\rm ss}$ is  $h\left(\{\wp_k\}\right)=-\sum_k \wp_k\log_2 \wp_k$.  This is lower bounded by the von Neumann entropy of $\rho_{\rm ss}$:   
\beq \label{boundonh}
h\left(\{\wp_k\}\right) \geq S(\rho_{\rm ss})=-\text{Tr}[\rho_{\rm  ss}\log_2 \rho_{\rm  ss}],
\eeq
with equality iff $\{\wp_k, \ket{v_k^e}\bra{v_k^e}\}$ is the diagonal ensemble. We showed in~\cite{KarWis11}, that some of the $K=2$ and $K=3$ ensembles for a qubit can have entropy $h$ much smaller than $1$. In this case one can track the state of the qubit with less than one bit on average, meaning that the state of $N$ identical qubits subject to the same independent monitoring can be tracked with $Nh \ll N$ bits. This is one reason why studying different PR ensembles for the same system is of great interest.  In the next section we explore the stability  of such adaptive monitoring.

\section{Jump Dynamics and Stability}
\label{sec:jumpDynamics}
 As explained above, the existence of a  PR ensemble $\{\wp_k, \ket{v_k^e}\bra{v_k^e}\}$ 
 ensures that if we start the system in state $\ket{v_k^e}$ and subject it to the adaptive monitoring determined by the parameters $\kappa_{jk}$ from Eq.~(\ref{jumpCond}), then the system will always jump between states $\ket{v_k^e}$ with $k=1,\ldots, K$ and throughout its  evolution will never leave  the PR ensemble. But what happens if the system  is  not initialized perfectly at the start of monitoring procedure? 

We address this question in this section.  We consider a general qubit system, whose evolution is governed by the master equation, Eq.~(\ref{me1}), with one jump operator, $\hat{c}$. We assume that this system is subject to adaptive monitoring that allows the system to jump between $K$ states in a cyclic manner. Since the resulting evolution is stochastic, we cannot say what will happen to each specific trajectory, but we can determine what happens to many different realizations on average. We show that on average any initial state subject to the adaptive  monitoring will eventually converge to the jumping between states $\ket{v_k^e}$ with $k=1,\ldots, K$ from PR ensemble. We first develop our results assuming that adaptive monitoring leaves the qubit jumping between $K=2$ states and then generalize to $K$-state cyclic jumps.

\subsection{Two-state jumping}
\label{subsec:2stJ}
For two-state jumping scenario, the system jumps between pure states $\ket{v_1^e}$ and $\ket{v_2^e}$. In principle, we can compute these states  using  \erfs{r1e}{r2e}. This approach tends to yield useful numerical answers, but extracting simple analytical expression is not easy.  An  alternative approach for identifying jumping states (for  the specific example of resonance fluorescence)  was  undertaken in~Refs.~\cite{WT99, KarWis10}. 
This approach  uses the  properties of transformations, \erfs{trans1}{trans2}, that leave master equation, eq.~(\ref{me1}), invariant, to explicitly construct the adaptive monitoring schemes that generates a PR ensemble 
with the desired number of elements. This is the approach we use in this section. 

The system we study in this paper is simple. It has only one jump operator and, therefore, the only degree of freedom for generating \erfs{trans1}{trans2} is the strength of the local oscillator, a  complex scalar $\mu$. For a given $\mu$, the effective Hamiltonian for the system is \cite{WisMil10}
\begin{equation}
\hat{H}(\mu)\equiv  \hat{H} -\frac{i}{2}\hat{c}^\dagger\hat{c}-i\mu^\ast\hat{c}-i\frac{|\mu|^2}{2} 
\label{eq:Hwithmu}
\end{equation}
and the jump operator is 
\begin{equation}
\hat{c}'(\mu)= \hat{c}+\mu.
\end{equation}
Note that although for convenience we are not using the ``eff'' subscript anymore, $\hat H(\mu)$ 
is the {\em non-Hermitian}  Hamiltonian $\hat H_{\rm eff}$ introduced earlier.

Under two-step adaptive monitoring, the signal from the system (a qubit) is mixed with the local oscillator with strength $\mu_1$ prior to the photon detection. Before the  photon  is detected the system undergoes smooth evolution governed by $\hat{H}_1=\hat{H}(\mu_1)$. Then the system experiences a jump governed by $\hat{s}_1=\hat{c}'(\mu_1)$  when  the photon is detected. At this point an experimentalist switches the strength of the local oscillator from $\mu_1$ to $\mu_2$ and the consequent smooth evolution is generated by the effective Hamiltonian $\hat{H}_2=\hat{H}(\mu_2)$ and the next jump is caused by $\hat{s}_2=\hat{c}'(\mu_2)$. As soon as  this  jump is detected, the experimentalist switches  the  strength of the local oscillator back to $\mu_1$ and the evolution cycle repeats itself.

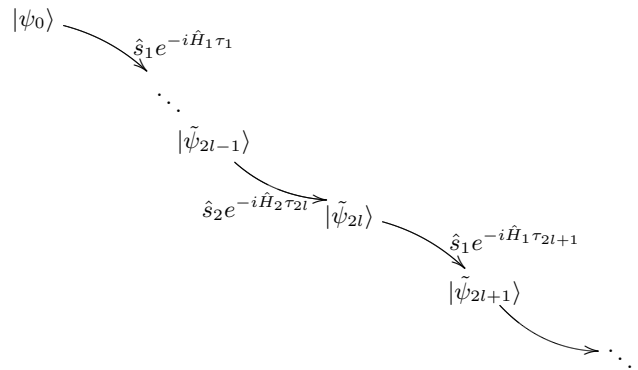
\begin{figure}
\begin{xy} 
{\ar@{->}@/^/ (0,24)*+{\ket{\psi_{0}}}; (18,14)*+{\ddots}} ?*!/_0mm/{\hspace{2.0cm}\hat{s}_1 e^{ -i\hat{H}_1 \tau_{1}}};
{\ar@{->}@/_/ (24,8)*+{ \ket{\tilde\psi_{2l-1}}}; (40,0)*+{}} ?*!/^0.75pc/{\hat{s}_2 e^{-i\hat{H}_2 \tau_{2l}}};
{\ar@{->}@/^/ (42,-2)*+{ \ket{\tilde\psi_{2l}}}; (60,-12)*+{ \ket{\tilde\psi_{2l+1}}}} ?*!/_0mm/{\hspace{2.4cm}\hat{s}_1 e^{-i\hat{H}_1 \tau_{2l+1}}};
{\ar@{->}@/_/ (62,-14)*+{}; (78,-20)*+{\ddots}}
\end{xy}
\caption{ A schematic of the conditioned state evolution under two-step adaptive monitoring. }
\label{fig:evol}
\end{figure}

We can  solve for the evolution of the system as following. Let the initial state of the system be $\ket{\psi_0}$. We assume that  the $n$th jump happens at time $t_n$  after which the system will be in state $\ket{\psi_n}$ and the waiting times between jumps are given by $\tau_n=t_n-t_{n-1}$. The evolution is stochastic and for each realization of such evolution jumps will happen at different times. The probability density function, $p(\vec{\tau})$, for waiting times $\vec{\tau}=(\tau_1, \ldots, \tau_{n})$ is given by 
\begin{equation}
p({\vec\tau})=||\ket{\tilde{\psi}_n}||^2
\label{eq:pdf}
\end{equation} 
Here the unnormalized state $\ket{\tilde{\psi}_{n}}$ is defined 
recursively by 
\begin{equation}
|\tilde{\psi}_{n}\rangle=\hat{s}_j e^{-i\hat{H}_j \tau_n}\ket{\tilde{\psi}_{n-1}}\text{ with } j=
\begin{cases} 1 & \mbox{for } n\mbox{ odd} \\ 2 & \mbox{for } n\mbox{ even}\end{cases}
\label{eq:unnormS}
\end{equation} 
 This construction implies that $\ket{\tilde{\psi}_{n-1}}$ implicitly depends on $\tau_1,\ldots, \tau_{n-1}$ and, consequently, state $\ket{\tilde{\psi}_{n}}$ depends on all prior waiting times $\vec{\tau}$. The diagram for this evolution is depicted in Fig.~\ref{fig:evol}
 At an arbitrary time $t=t_{n-1}+s$, with $0 \leq s < \tau_{n}$, for some $n$, 
the state of the system is given by  
\begin{equation}
\ket{\psi(t)}=\frac{e^{ -i\hat{H}_1 s}\ket{\tilde{\psi}_{n-1}}}{||e^{-i\hat{H}_1 s}\ket{\tilde{\psi}_{n-1}}||} \text{ with } j=
\begin{cases} 1 & \mbox{for } n\mbox{ odd} \\ 2 & \mbox{for } n\mbox{ even}\end{cases}  .
\label{eq:bwtjumpev}
\end{equation}

We can force the system evolution to be restricted to just two states, $\ket{v_1^e}$ and $\ket{v_2^e}$ if we require that the state $\ket{v_1^e}$ is an eigenstate of $\hat{H}_1$, the state $\ket{v_2^e}$ is an eigenstate of $\hat{H}_2$, the jump operator $\hat{s}_1$ maps $\ket{v_1^e}$ to $\ket{v_2^e}$ and $\hat{s}_2$ performs the reverse action.  A consequence of these assumptions is that after two jumps the state $\ket{v_1^e}$ should return back to itself, i.e.
\begin{equation}
\hat{s}_1 \hat{s}_2 \ket{v_1^e}\equiv [\mu_1\mu_2 +(\mu_1+\mu_2)\hat{c} +\hat{c}^2]\ket{v_1^e} \propto \ket{v_1^e}
\label{eq:jumpOp}
\end{equation}
Without loss of generality, we can assume that the jump operator $\hat{c}$ is traceless and this means that the square of this operator is proportional to the identity, $\hat{c}^2\propto \hat{I}$.  Thus, Eq.~(\ref{eq:jumpOp}) can happen only in two ways:  either (i)  $\ket{v_1^e}$ is an eigenstate of  $\hat{c}$; or (ii) $\mu_1+\mu_2=0$.  Case  (i) holds iff $\ket{v_1^e}$ is an eigenstate of $\hat{s}_1$, which would prevent any jumping from happening. Since  case (i) is ruled out,  case (ii) must hold. Thus, $\mu_1 = -\mu_2$  and the full cycle jump operator $\hat{s}_1\hat{s}_2$ is proportional to the identity, which is a very useful fact for the discussion later on. 

We also comment here how \erfs{eq:pdf}{eq:bwtjumpev} can be used for simulation of a piecewise deterministic process and, in particular, for two-state jumping. We start the system in the normalized state $\ket{\psi_0}$. We know that after $n$th jump the normalized state of the system is given by Eq.~(\ref{eq:bwtjumpev}) with $s=0$. We determine a random waiting time $\tau_n$ according to the cumulative waiting time distribution \cite{BrePet02}
\begin{equation}
F(\tau_n)=1-||e^{-i \hat{H}_j \tau_n}\ket{\psi(\tau_{n-1})}||^2
\label{cdf}
\end{equation}
with $j=1$ if $n$ is odd and $j=2$ if $n$ is even. This is done as follows. We draw a random number $\eta$ from the uniform distribution over the interval $[0,\,1]$ and  solve $\eta=1-F(\tau_n)$ for $\tau_n$. Between the jumps the state of the system is given by Eq.~(\ref{eq:bwtjumpev}). The evolution depicted in Figs.~\ref{unEvol} and \ref{typEvol} is generated using this procedure. 

The cumulative waiting time distribution, $F(\tau_n)$ can be, of course, related back to the probability density $p(\vec{\tau})$. Using the induction step
\begin{equation}
p(\tau_n)=p(\tau_n|\tau_{n-1}) \,p(\tau_{n-1})=\frac{d}{d\tau_n} F(\tau_n) p(\tau_{n-1})
\end{equation}
together with the fact that 
\begin{align}
\frac{d}{d\tau_n} F(\tau_n) &=
-\frac{d}{d\tau_n}  ||e^{-i\hat{H}_j \tau_n}\ket{\psi(\tau_{n-1})}||^2 \notag \\
&=\langle \psi(\tau_{n-1})|e^{i\hat{H}_j^\dagger \tau_n}\hat{s}_j^\dagger \hat{s}_j e^{-i\hat{H}_j \tau_n}|\psi(\tau_{n-1})\rangle \notag \\
&=\frac{||\ket{\tilde{\psi}_n}||^2}{||\ket{\tilde{\psi}_{n-1}}||^2},
\label{eq:probDen}
\end{align}
we can deduce Eq.~(\ref{eq:pdf}).

\subsection{Mean-square stability}
\label{subsec:cstab}
We now know how to describe the state of the system subject to two-state monitoring  and we know the probability density for jumps occurring at times $t_1, \dots, t_n$ with waiting times $\vec{\tau}=(\tau_1, \ldots, \tau_n)$. The resulting evolution is a stochastic process, as we don't know when the jumps   occur.  Thus 
it is sensible to look at what happens to the system on average, over many different realizations,  by considering the mean-square stability. 

The mean-square  fidelity between the desired system state  and the actual system state immediately following the $n$th jump, starting from initial state $\ket{\psi_0}$ is 
\begin{align}
\Big\langle|\langle v_j^e|\psi_{n}\rangle|^2\Big\rangle&=\int d^n\vec{\tau} \, p({\vec\tau}) |\langle v_j^e|\psi_{n}\rangle|^2 \notag \\ &=\int d^n{\vec\tau}\, ||\ket{\tilde{\psi}_n}||^2 \frac{|\langle v_j^e|\tilde{\psi}_{n}\rangle|^2}{||\ket{\tilde{\psi}_n}||^2} \notag \\
&=\int d^n{\vec\tau}\, |\langle v_j^e|\tilde{\psi}_{n}\rangle|^2
\end{align}
with $j=1$ if $n$ is odd and $j=2$ if $n$ is even. Thus this quantity is easy to compute. At the same time, it is easy to see that if 
\begin{equation}
\lim_{n\to \infty} \Big\langle|\langle v_j^e|\psi_{n}\rangle|^2 \Big\rangle= 1\text{ with } j=\begin{cases} 1 & \mbox{if } n\mbox{ is even} \\ 2 & \mbox{if } n\mbox{ is odd} \end{cases}
\end{equation}
the system will  converge on average to a perfect jumping scenario.  We call this \emph{mean-square stability}. 

We now investigate under what conditions the system has mean-square stability. 
To do this, we need a way to evaluate the dependence of $\ket{\tilde{\psi}_n}$  on the jump times, ${ \vec\tau}$. We let $\ket{v_j^e}$ and $\ket{v_j^{{o}}}$ be the eigenstates of $i\hat{H}_j$ with respective eigenvalues $\lambda_j^e$ and $\lambda_j^{{ o}}$, where states $\ket{v_j^e}$ are  part of the PR {\em ensemble},  
and states $\ket{v_j^{{ o}}}$ are {\em other} states, not part of the ensemble.  It is easiest to compute the time dependence of $\ket{\tilde{\psi}_n}$ if we decompose this state in terms of eigenstates of $\hat{H}_1$ for even $n$ (i.e., $\ket{v_1^e}$ and $\ket{v_1^{{ o}}}$) and in terms of eigenstates of  $\hat{H}_2$ for odd  $n$ (i.e., $\ket{v_2^e}$ and $\ket{v_2^{{ o}}}$).

Thus we let the state of the system after ${2l}$ and ${2l}+1$ jumps be  respectively 
\begin{align}
\ket{\tilde{\psi}_{{ 2l}}}&=\alpha_{{ 2l}}\ket{v_1^e}+\beta_{{ 2l}} \ket{v_1^{{ o}}} \label{tev1}\\
\ket{\tilde{\psi}_{{ 2l}+1}}&=\alpha_{{ 2l}+1}\ket{v_2^e}+\beta_{{ 2l}+1} \ket{v_2^{{ o}}}.
\label{tev2}
\end{align}
The dependence on the jump times is hidden in the coefficients $\alpha_{{ 2l}}$, $\alpha_{{ 2l}+1}$, $\beta_{{ 2l}}$,  and $\beta_{{ 2l}+1}$, which we determine by establishing the relationship between them and their dependence on $\alpha_0$ and $\beta_0$.  To do this, we first define the action of jump operators $\hat{s}_1$ and $\hat{s}_2$ is on the basis states $\{\ket{v_1^e},\,\ket{v_1^{{ o}}}\}$ and $\{\ket{v_2^e},\,\ket{v_2^{{ o}}}\}$: 
\begin{align}
\hat{s}_1 \ket{v_1^e}&=Q^1_{11} \ket{v_2^e} \notag \\ 
\hat{s}_1 \ket{v_1^{{ o}}} &=Q^1_{21} \ket{v_2^e} +Q^1_{22}\ket{v_2^{{ o}}}\notag \\
\hat{s}_2 \ket{v_2^e}&=Q^2_{11} \ket{v_1^e} \notag \\ 
\hat{s}_2 \ket{v_2^{{ o}}}& =Q^2_{21} \ket{v_1^e} +Q^2_{22}\ket{v_1^{{ o}}} .
\label{srel}
\end{align}
 The scalar elements $Q^1_{jk}$ and $Q^2_{jk}$ can be determined once all the parameters of the system, $\hat{H}$,  $\hat{c}$, and $\mu_1$, are specified. Eqs.~(\ref{srel}) simply state that jumping operators ($\hat{s}_1$ and $\hat{s}_2$) map jumping states ($\hat{v}_1^e$ and $\hat{v}_2^e$) to each other and the mapping for the other basis states is quite general.

Using \erfs{tev1}{srel} and propagation between states given in Eq.~(\ref{eq:unnormS}), we can derive 
\begin{align}
\beta_{{ 2l}+1}&=\beta_{{ 2l}}e^{-\lambda_1^{{ o}} { \tau}_{{ 2l}+1}} Q^1_{22}\label{betarel1}\\
\beta_{{ 2l}}&=\beta_{{ 2l}-1}e^{-\lambda_2^{{ o}} { \tau}_{{ 2l}}}Q^2_{22}.\label{betarel2}
\end{align}
Now we can show by induction that 
\begin{align}
\beta_{{ 2l}}&=\beta_0 (Q^1_{22} Q^2_{22})^l\prod_{j=1}^l e^{-\lambda_1^{{ o}} { \tau}_{2j-1}}e^{-\lambda_2^{{ o}} { \tau}_{2j}} \label{beta2l} \\
\beta_{{ 2l}+1}&=\beta_0 Q^1_{22}(Q^1_{22} Q^2_{22})^l\prod_{j=0}^l e^{-\lambda_2^{{ o}} { \tau}_{2j}} e^{-\lambda_1^{{ o}} { \tau}_{2j+1}}.
\label{beta2lp1}
\end{align}

This information is sufficient to derive the mean-square fidelity $\big\langle |\langle v_j^e|\psi_n\rangle|^2\Big \rangle$,  using the following useful identity 
\begin{equation}
|\langle v_j^e|\tilde{\psi}_n\rangle|^2=||\ket{\tilde{\psi}_n}||^2-|\beta_n|^2(1-|O_j|^2),
\label{eq:Uidentity}
\end{equation}
where $j=1$ if $n$ is even and $j=2$ if $n$ is odd and $O_j=\langle v_j^e|v_j^{{o}}\rangle$ is the overlap function. 
Now  calculating the mean-square fidelity becomes easy. According to Eq.~(\ref{eq:pdf}), $||\ket{\tilde{\psi}_n}||^2$ is the probability density for jumps occurring with waiting times ${\vec\tau}=({ \tau}_1, \ldots, { \tau}_n)$  and thus when integrated with respect to $d^n{\vec\tau}$ yields one. Thus
\begin{equation}
\Big\langle|\langle v_j^e|\psi_n\rangle|^2\Big\rangle=1- (1-|O_j|^2) \int  d^n\vec\tau  |\beta_n|^2, 
\label{eq:msOvS1}
\end{equation}
with $j=1$ if $n$ is even and $j=2$ if $n$ is odd. 
Using Eq.~(\ref{beta2l}), we learn
\begin{equation}
\int  d^{2l}\vec\tau  |\beta_{2l}|^2 = |\beta_0|^2 \left(\frac{|Q^1_{22} Q^2_{22}|^2}{4 {\rm  Re}\lambda_1^{{ o}} {\rm  Re}\lambda_2^{{ o}}}\right)^l\equiv |\beta_0|^2 C^l.
\label{eq:abeta2l}
\end{equation}
Thus the mean-square fidelity is
\begin{equation}
\Big\langle|\langle v_1^e|\psi_{{2l}}\rangle|^2\Big\rangle=1-|\beta_0|^2 (1-|O_1|^2)C^l
\label{eq:msF}
\end{equation}
and
\begin{equation} \label{eq:meanC}
\Big\langle|\langle v_2^e|\psi_{ 2l+1}\rangle|^2\Big\rangle=1-|Q^1_{22}\beta_0|^2 (1-|O_2|^2)C^l.
\end{equation}
Thus coefficient $C$ determines the convergence rate to the perfect two-state jumping.

The absolute value of fidelity is a number between $0$ and $1$ and its average with respect to all possible waiting times $\vec{\tau}$ is still between $0$ and $1$. Thus Eq.~(\ref{eq:msF}) implies that $C \leq 1$. 
Thus unless $C=1$, \erf{eq:meanC} implies that the  mean-square fidelity always converges to 1 and the system has mean-square stability.

We now determine under what conditions parameter $C$ is strictly less than one. To answer this question, we need one more calculation, which is performed in Appendix~A and establishes that 
$\xfrac{|Q^1_{11} Q^2_{11}|^2}{4 {\rm  Re}\lambda_1^e {\rm  Re}\lambda_2^e}=1$ and that $Q^1_{11} Q^2_{11}=Q^1_{22} Q^2_{22}$. 
Thus 
\beq
C=\frac{{\rm  Re}\lambda_1^e {\rm  Re}\lambda_2^e}{{\rm  Re}\lambda_1^{{ o}} {\rm  Re}\lambda_2^{{ o}}}.
\end{equation}
In other words,   the system will have mean-square stability iff 
the geometric mean of the real part of the eigenvalues of the states {\em in} the ensemble is smaller than the geometric mean of the real part of the eigenvalues of the states {\em not in} the ensemble. This makes sense, as the 
real parts of these eigenvalues gives the rate at which the amplitudes of their respective states decay during the between-jump stages of the evolution, 
as we will explore in Sec.~\ref{Sec:stages}. Thus, the smaller they are, 
the more stable the respective states. As we will see in Sec.~\ref{Sec:Fluorostable2}, the  condition for mean-square stability 
can be satisfied even though every second stage is {\em unstable}.

\subsection{$K$-state jumping}
\label{Sec:Kjump}
We now explain how our results can be generalized to cyclic $K$-state jumping, 
as first studied in Ref.~\cite{KarWis11} for $K>2$. 
 We simply let the index $j$ for expressions in sections~\ref{subsec:2stJ} and \ref{subsec:cstab} range over $\{1, \ldots, K\}$ instead of just taking values $1$ and $2$. All considerations still hold and we can derive Eq.~(\ref{eq:msOvS1}) as before, but the index $j$ in this expression is now determined according to $j= (n \mod K)  +1$. We also need to reevaluate $\int d^n\vec\tau  |\beta_n|^2$. The unnormalized state for the $j$th step in cycle after $n$ repeats is 
\begin{equation}
 \ket{\tilde{\psi}_{N}}=\alpha_N\ket{v_{j+1}^{{ e}}}+\beta_N \ket{v_{j+1}^{{ o}}}
\end{equation}
with $N=nK+j$ and $j+1$ stands for $(j+ 1) \mod K $. Equation~(\ref{srel}) becomes
\begin{align}
\hat{s}_j \ket{v_j^e}&=Q^j_{11} \ket{v_{j+1}^{{ e}}} \notag \\ 
\hat{s}_j \ket{v_j^{{ o}}} &=Q^j_{21} \ket{v_{j+1}^{{ e}}} +Q^j_{22}\ket{v_{j+1}^{{ o}}}\label{eq:lstep}
\end{align}
and expression for $\beta_{n K}$ becomes
\begin{equation}
\beta_{n K}=\beta_0 (Q^1_{22} \cdots Q^K_{22})^n\prod_{j=0}^{n-1} e^{-\lambda_1^{{ o}} { \tau}_{jK+1}}\cdots e^{-\lambda_K^{{ o}} { \tau}_{(j+1)K}}
\end{equation}
and thus 
\begin{equation}
\int d^{nK}\vec\tau|\beta_{n K}|^2 =|\beta_0|^2 \left(\frac{Q^1_{22}\cdots Q^K_{22}}{2^n {\rm  Re}\lambda_1^{{ o}} \cdots {\rm  Re}\lambda_K^{{ o}}}\right)^n.
\end{equation}
Thus parameter $C$ that determines stability is redefined to be
\begin{equation}
C=\frac{Q^1_{22}\cdots Q^K_{22}}{2^n {\rm  Re}\lambda_1^{{ o}} \cdots {\rm  Re}\lambda_K^{{ o}}}.
\end{equation}
Just as before we can  show  that $C$ can be re-expressed as 
\beq \label{eq:KstateC}
C = \frac{{\rm  Re}\lambda_1^e \cdots {\rm  Re}\lambda_K^e}{{\rm  Re}\lambda_1^{{ o}} \cdots {\rm  Re}\lambda_K^{{ o}}},
\eeq
so that  $C<1$ whenever ${\rm  Re}\lambda_1^{{ o}} \cdots {\rm  Re}\lambda_K^{{ o}}>{\rm  Re}\lambda_1^e \cdots {\rm  Re}\lambda_K^e$. The argument proceeds in the same way as for $2$-state jumping,  and the details appear in Appendix A. 

We conjecture that the fact that $\ket{v_j^e}$ form a PR ensemble for $\rho_{\rm  ss}$, the unique steady state, ensures that  $\prod_{ k=1}^{ K} {\rm  Re}\lambda_k^{{ o}}>\prod_{ k=1}^{ K} {\rm  Re}\lambda_k^e$. Even more generally, we conjecture that every finite PR ensemble is stable.

\subsection{ Stability of an individual trajectory} \label{Sec:stages}

It is important to note that mean square stability  does not imply that in an individual trajectory there will be 
 monotonic convergence of the 
 system towards the desired states.  The fidelity could decrease  between the jumps, and/or could decrease
 due to a jump.  We   begin with the first issue. 
 
 We say that the system is \emph{piecewise stable} if $|\langle v_j^e|\psi(t)\rangle|^2$ is a monotonically increasing function of $t$ during all stages 
$t_{n-1} \leq t < t_n$,  where $j=1$ if $n$ is odd and $j=2$ if $n$ is even. 
Otherwise, if the fidelity with the desired state decreases between jumps,  
 this constitutes an \emph{unstable stage} in the evolution. As we will show, this happens only if $n$ even or $n$ odd, 
 not both. That is, stable and unstable stages will alternate. 
 
 The stability of the  evolution between jumps  is determined by the relationship between eigenvalues $\lambda_j^e$ and $\lambda_j^{{ o}}$. 
Let us assume without loss of generality that the state of the system right after the jump is
\begin{equation}
\ket{\psi_0}=\alpha_0\ket{v_j^e}+\beta_0 \ket{v_j^{{ o}}}.
\end{equation}
Then the unnormalized state of the system at time $\tau$ after the jump, but before the next one, is
\begin{equation}
\ket{\tilde{\psi}(\tau)}=\alpha_0 e^{-\lambda_j^e \tau}\ket{v_j^e}+\beta_0 e^{-\lambda_j^{{ o}} \tau}\ket{v_j^{{ o}}}.
\end{equation}
We now want to know that happens to the fidelity between $\ket{\psi(\tau)}$ (normalized state of the system between jumps) and the ideal jumping state $\ket{v_j^e}$: 
\begin{equation}
|\langle v_j^e|\psi(\tau)\rangle|^2=\frac{|\langle v_j^e|\tilde{\psi}(\tau)\rangle|^2}{||\ket{\tilde{\psi}(\tau)}||^2}.
\end{equation}
Using Eq.~(\ref{eq:Uidentity}), we can compute  this fidelity to be 
\begin{equation}
|\langle v_j^e|\psi (\tau)\rangle|^2=1-\frac{|\beta_0|^2(1-|O_j|^2)}{|\alpha_0 e^{-(\lambda_j^e-\lambda_j^{{ o}})\tau}\ket{v_j^e} + \beta_0 \ket{v_j^{{ o}}}|^2}.
\end{equation}
Thus we can see that if ${\rm Re}\lambda_j^e<{\rm Re}\lambda_j^{{o}}$ then in the long time limit fidelity  approaches one. That is,  between jumps  the system will converge to the  desired ensemble state with exponential rate and  the  fidelity is monotonically increasing to the maximum value  of one. On the other hand, if ${\rm Re}\lambda_j^e>{\rm Re}\lambda_j^{{ o}}$, then the system state converges towards the non-ensemble eigenstate, and the   
fidelity converges towards $|O_j|^{2}$, which is a quantity less than one. Thus the fidelity may  decrease during such a stage, and we call this an unstable stage in the evolution. This is illustrated for the case of resonance fluorescence in \srf{Sec:Fluorostable2}. 

We now address the issue of whether the jump itself increases or decrease fidelity. 
We again introduce some notation. Let $\ket{\tilde{\phi}_1}=\alpha_1 \ket{v_j^e}+\beta_1 \ket{v_j^{{ o}}}$ be the state of the system right before the jump and $\ket{\tilde{\phi}_2}=\alpha_2 \ket{v_k^e}+\beta_2 \ket{v_k^{{ o}}}$ be the unnormalized state of the system right after the jump, where $k= (j+1) \mod K$.  We also know that $\ket{\tilde{\phi}_2}=\hat{s}_j\ket{\tilde{\phi}_1}$. We want to compare the fidelity right before and immediately after the jump and we let 
\begin{equation}
F_1=\frac{|\langle v_j^e|\tilde{\phi}_1\rangle|^2}{||\ket{\tilde{\phi}_1}||^2} \text{ and }F_2=\frac{|\langle v_k^e|\tilde{\phi}_2\rangle|^2}{||\ket{\tilde{\phi}_2}||^2}.
\end{equation}
We evaluate $ |\langle v_j^e|\tilde{\phi}_1\rangle|^2$ and $ |\langle v_k^e|\tilde{\phi}_2\rangle|^2$ using Eq.~(\ref{eq:Uidentity}). Then fidelity will decrease after the jump, $F_1<F_2$, iff 
\begin{equation}
\frac{||\ket{\tilde{\phi}_2}||^2}{||\ket{\tilde{\phi}_1}||^2}<\frac{|\beta_2|^2}{|\beta_1|^2}\frac{1-|O_k|^2}{1-|O_j|^2} \equiv B.
\label{eq:fjcond}
\end{equation}

We can easily determine tight bounds for $\xfrac{||\ket{\tilde{\phi}_2}||^2}{||\ket{\tilde{\phi}_1}||^2}$ by observing that   
\begin{equation}
\lambda_{\rm  min}||\ket{\tilde{\phi}_1}||^2  \leq ||\ket{\tilde{\phi}_2}||^2 = \langle  \tilde{\phi}_1 |\hat{s}_j^\dagger \hat{s}_j |\tilde{\phi_1}\rangle  \leq  \lambda_{\rm max}||\ket{\tilde{\phi}_1}||^2,
\label{eq:lubound}
\end{equation}
where $\lambda_{\rm  min}$ and $\lambda_{\rm max}$
are the smallest and the largest eigenvalues of the operator $\hat{s}_j^\dagger\hat{s}_j$. Thus the  lower bound of $\lambda_{\rm min}$, or upper bound of $\lambda_{\rm max}$,  is attained if $\ket{\tilde{\phi}_1}$  is the corresponding eigenstate of $\hat{s}_j^\dagger\hat{s}_j$. 
If $\lambda_{\rm min} < B$, it is possible to observe jumps that decrease the fidelity. 
Again, we illustrate this for the resonance fluorescence example in \srf{Sec:Fluorostable2}. Note however that for mean-square stable evolution, in the long time limit, $\ket{\phi_1}$ converges to $\ket{v_j^e}$, and so $F_1$ and $F_2$ also converge to one. 
Thus we expect that as the system converges towards the ideal ensemble states, observing jumps that decrease fidelity becomes less and less likely.

\section{Resonance fluorescence} 
\label{sec:Resfluor}
In this section we apply our results for stability in the mean-square sense, and for individual trajectories, to a specific 
physical example.  In particular, we show that there are PR ensembles  which are mean-square stable, but whose cycles are composed from stable and unstable stages. Others are comprised of only stable stages, and so the fidelity is piecewise monotonically increasing. It is only piecewise because this example also illustrates  that jumps can decrease the fidelity.  

The example we consider is resonance fluorescence. We consider a qubit (for example, a two-level atom) with basis states $\ket{0}$ and $\ket{1}$, with a transition frequency $\omega_0$. We assume that qubit is coupled to the continuum of electromagnetic radiation and, therefore, decays to $\ket{0}$ at rate $\gamma$. At the same time, it is driven by a classical field 
oscillating at frequency $\omega_0$. The strength of the driving is quantified by the  Rabi frequency $\Omega$. 
In the interaction frame \cite{WisMil10} with respect to the atomic transition frequency $\omega_0$, 
the evolution of the qubit is given by the master equation of the form of Eq.~(\ref{me1}) with $\hat H= (\Omega/2) \hat{\sigma}_x$ and one jump operator $\hat c = \sqrt{\gamma}\hat{\sigma}$, where $\hat{\sigma}=\ket{0}\bra{1}$ and $\hat{\sigma}_x=\hat{\sigma}+\hat{\sigma}^\dagger$. Then matrix $A$ and $\vec{b}$ in Bloch vector equation, Eq.~(\ref{bve}), are
\begin{equation}
 A=\begin{pmatrix} 
    -\gamma/2& 0&0\\
    0&-\gamma/2&-\Omega\\
    0&\Omega&-\gamma
   \end{pmatrix}\,\,\, \text{ and }
\vec{b}=\begin{pmatrix}
         0\\ 0\\ \gamma
        \end{pmatrix}. 
\end{equation}
The steady state $\vec{r}_{\rm ss}= (0, 2\gamma\Omega,-\gamma^2)^T/\ro{\gamma^2+2\Omega^2}$, is a mixed state for $\Omega\neq0$. 
The unnormalized eigenvectors of $A$ are $\vec{u}_1=(1,\,0\,0)^T$ and $\vec{u}_\pm=(0, \gamma\pm\sqrt{\gamma^2-16 \Omega^2}, 4 \Omega)^T$. For $\varepsilon\equiv\Omega/\gamma$ and $|\varepsilon|<1/4$ all three eigenvectors of $A$ are real, while for $|\varepsilon|>1/4$ only $\vec{u}_1$ is real. Note that $\varepsilon$ here is different from $\epsilon = (\Omega/\gamma)^2$ in Ref.~\cite{KarWis11}.

\subsection{Two-state jumping} \label{Sec:Fluoro2}

We now analyze stability properties for two-state jumping in this system and we do this with the method described in \srf{subsec:cstab} and in~Refs.~\cite{WT99, KarWis10}, i.e., we know that PR ensemble is constructed from the eigenvectors of $\hat H(\mu)$, or equivalently,  of $i\hat{H}(\mu)$, where  $\hat H(\mu)$ is the effective (non-Hermitian) Hamiltonian  given by Eq.~(\ref{eq:Hwithmu}). 

\begin{figure}
\subfigure[]{
 \includegraphics[width=3.75cm, 
trim=10 10 10 10]{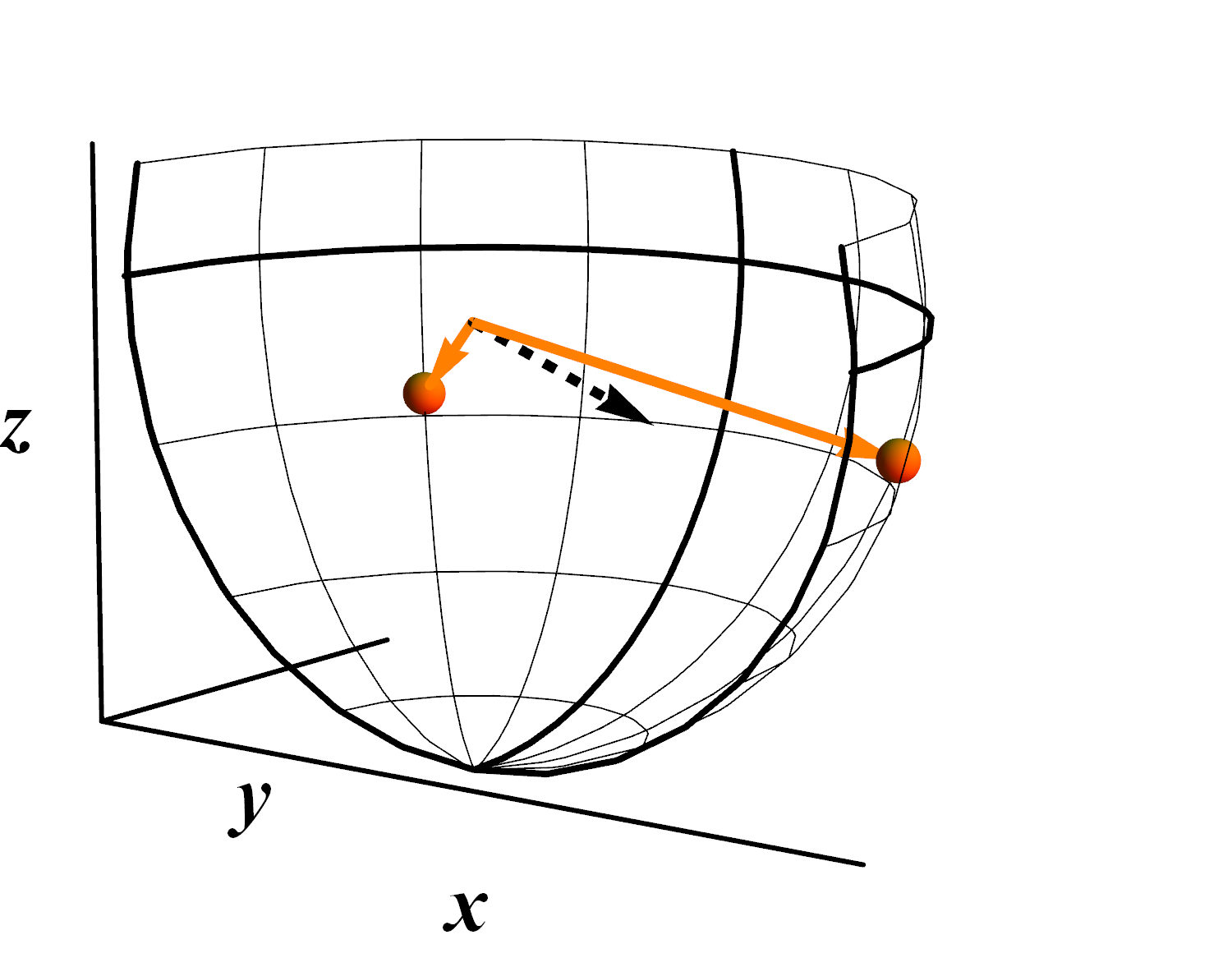}
}\hspace{-2.5 mm}%
\subfigure[]{
 \includegraphics[width=3.75cm, 
trim=10 10 10 10]{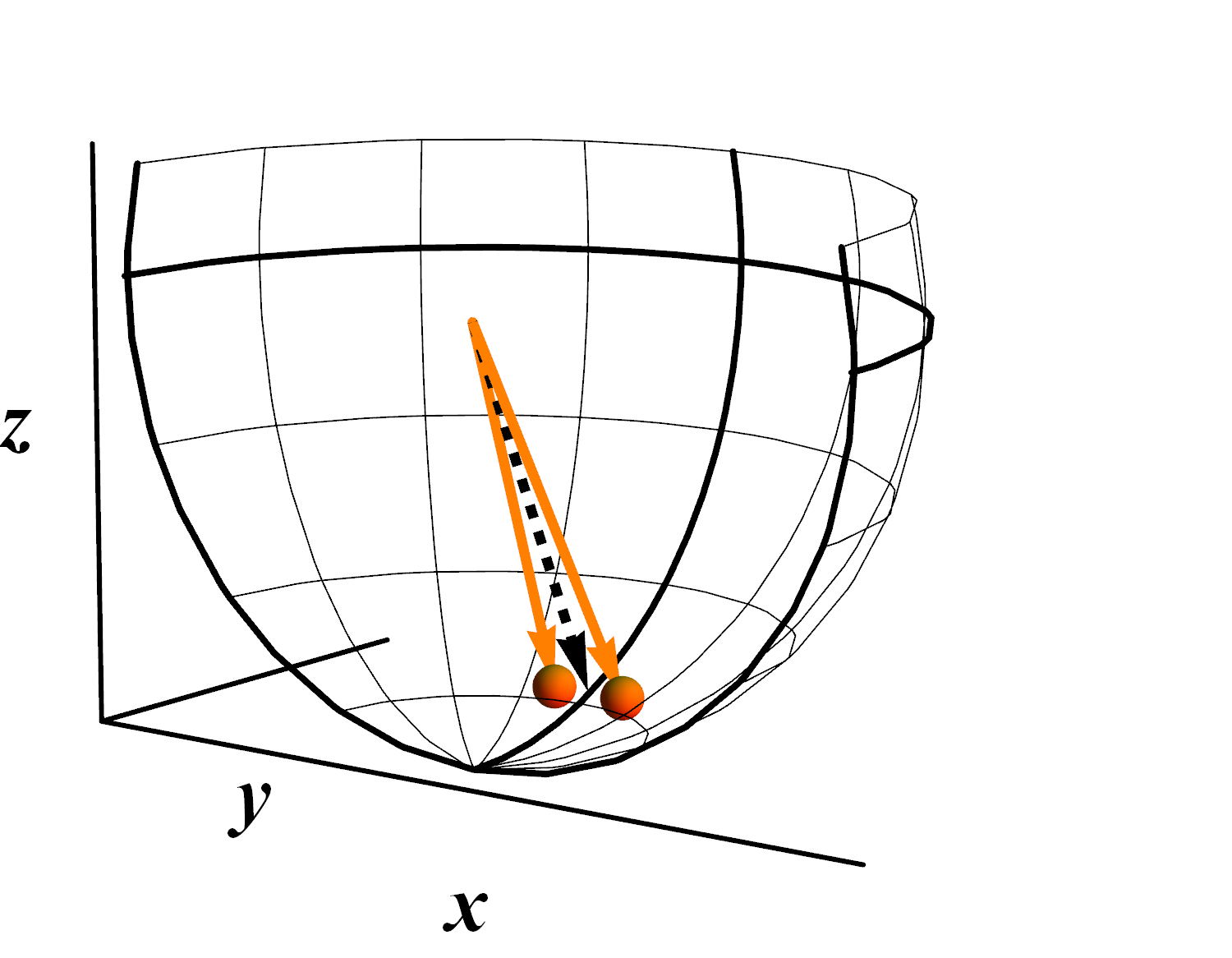}
}\vspace{-2 mm}
\subfigure[]{
 \includegraphics[width=3.75cm, 
trim=10 10 10 10]{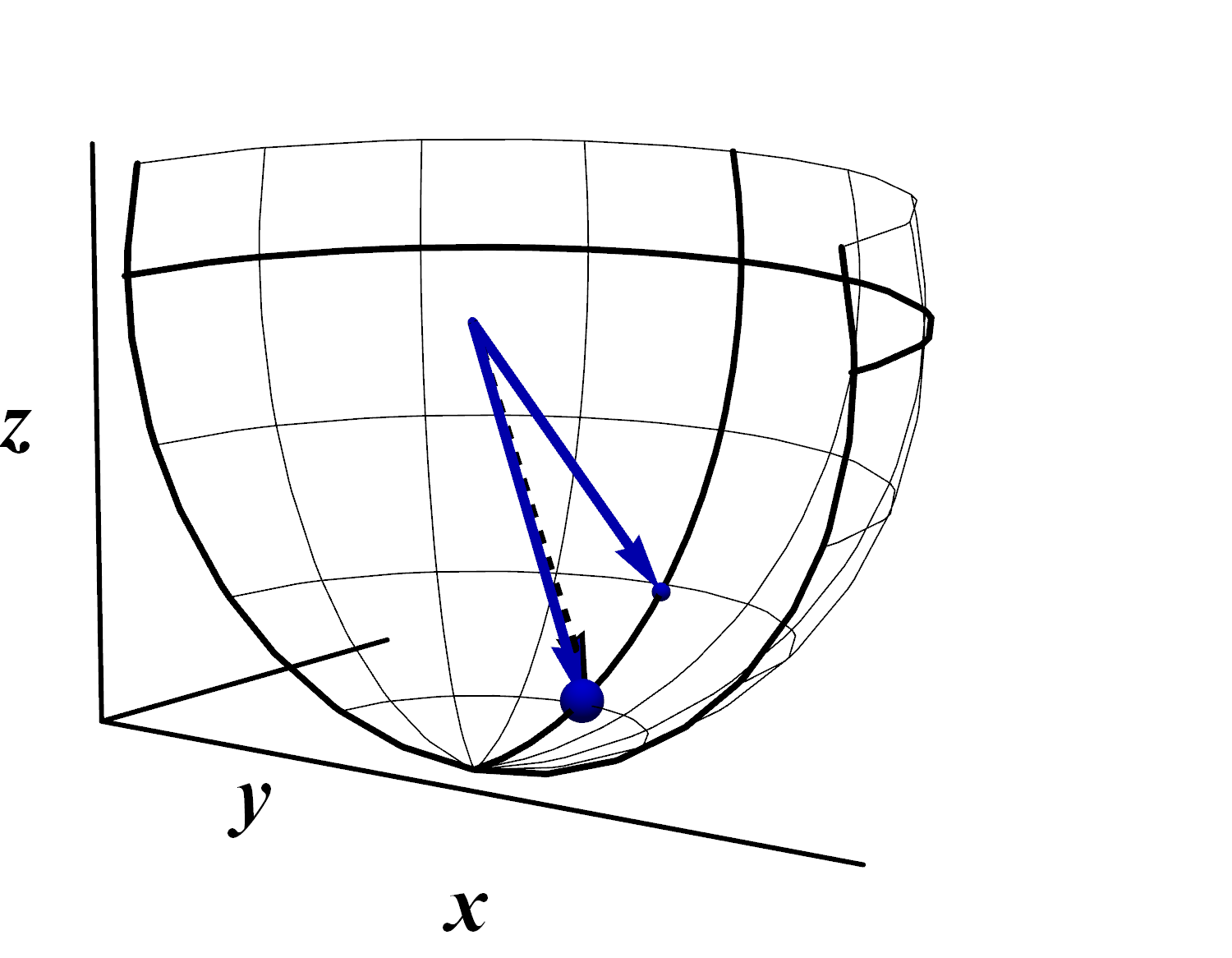}
}\hspace{-2.5 mm}%
\subfigure[]{
 \includegraphics[width=3.75cm, 
trim=10 10 10 10]{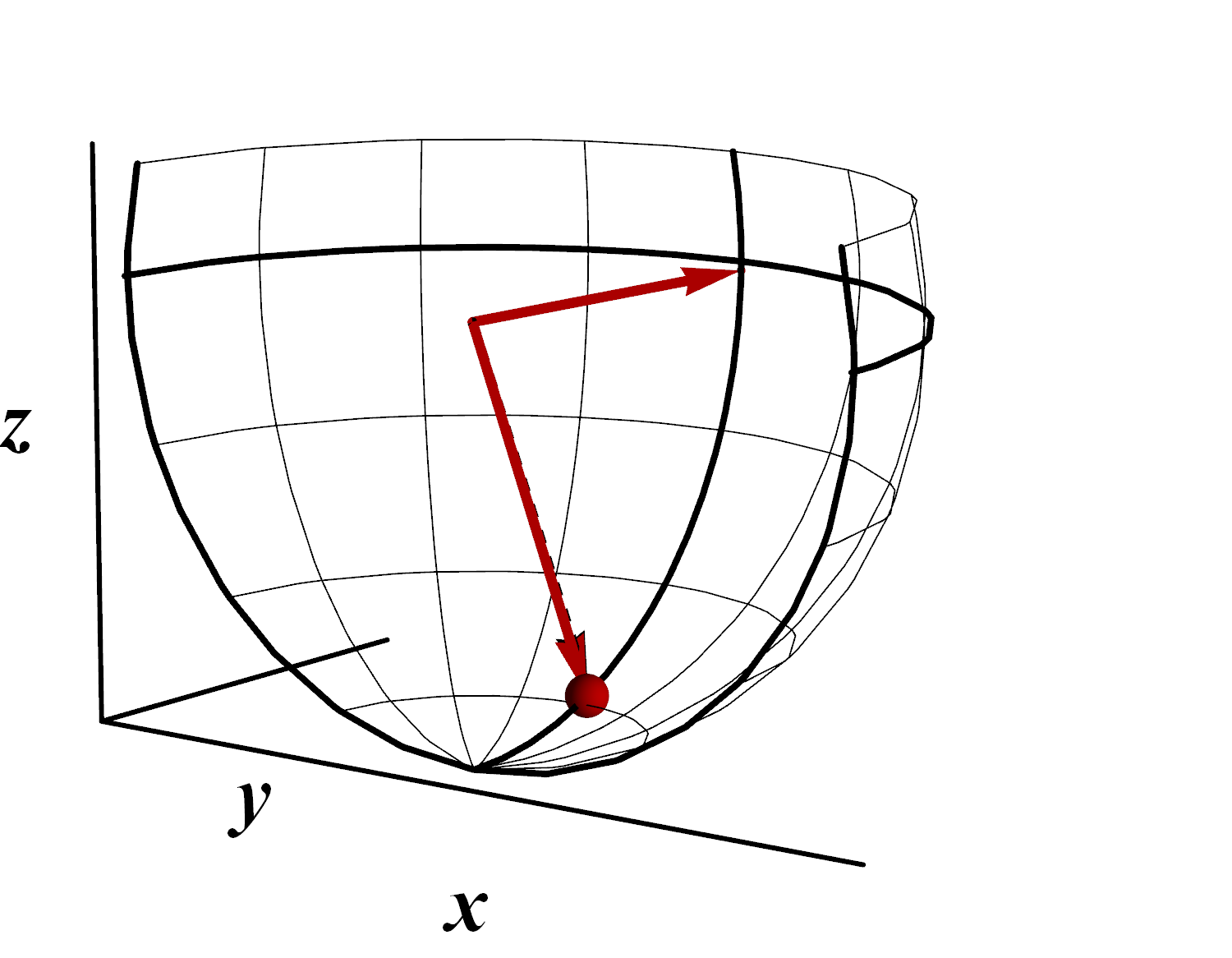}
}\vspace{-2mm}
\caption{(Color online.) Solid arrows show Bloch vectors for 2-state jumping. The volume of the sphere at the tip of each arrow represents the probability that the qubit occupies the corresponding pure state.  The dashed arrow is $\vec{r}_{\rm ss}$. Solutions in (a) and (b) arise from $\vec{u}_1$ with $\varepsilon=1$ and $\varepsilon=0.23$, respectively. This is the solution that exists for all $\varepsilon$. Solutions depicted in (c) and (d) exist only for $\varepsilon\leq 0.25$ and shown here for $\varepsilon=0.23$. The solution in (c) is generated by $\vec{u}_+$ and that in (d) by $\vec{u}_-$.}
\label{2sj}
\vspace{-5mm}
\end{figure}

For resonance fluorescence, the operator $-i\hat{H}(\mu)$  has eigenvectors given by
\begin{equation}
 \ket{v_\pm(\mu)}=\varepsilon \ket{1}+\left(\pm \sqrt{\varepsilon^2-\tfrac{1}{4}-2i\varepsilon \mu^\ast}+\tfrac{i}{2}\right)\ket{0}.
\end{equation}
The corresponding eigenvalues are 
\begin{equation} \label{eigmu}
\lambda_\pm(\mu)=\frac{1+2|\mu|^2}{4}\pm \frac{i}{2} \sqrt{\varepsilon^2-\tfrac{1}{4}-2i\varepsilon \mu^\ast},
\end{equation} 
where here, and in the remainder of the paper, we have set $\gamma=1$ for simplicity. 
Whether $\ket{v_+(\mu)}$ or $\ket{v_-(\mu)}$ is part of the PR ensemble depends on a particular value of $\mu_1$.
We know already from \srf{subsec:2stJ} that $\mu_2=-\mu_1$, so without loss of generality we can choose $\mu_1$
to have a non-negative real part.  
 Once we impose additional conditions from \srf{3stateJump} associated with jump operators $\hat{s}_1$ and $\hat{s}_2$, we learn that $\mu_1$ only assumes values from the set  
$\left\{\xfrac{1}{2}, \nu_+, \nu_-\right\}$, 
where 
\beq
\nu_\pm = i \frac{\sqrt{1\pm \sqrt{1-16\varepsilon^2}}}{2\sqrt{2}}.
\end{equation} 
The last two values in the set contribute to solutions only for $\varepsilon\leq1/4$.  That is, 
one only obtains a PR ensemble for values of $\mu_1$ that are either real or purely imaginary. 
Detailed derivation of these conditions can be found in Refs.~\cite{WT99, KarWis10}, 
and the Bloch vectors for all three ensembles are shown in  Fig.~\ref{2sj}. 


For $\mu_1=1/2$, the jumping states (in \erfs{r1e}{r2e}) are spanned from eigenvector $\vec{u}_1$ and, as explained in \crf{KarWis11}, the system spends equal amount of time in each state. Equivalently, the probability of occupying each state is $1/2$ and Shannon entropy is 1. Thus one bit is sufficient to track the state of the system. As shown in Fig.~\ref{2sj}(a)-(b), the solution is symmetric with respect to $x=0$ plane and as $\varepsilon$ increases jumping states move away from each other and away from the bottom of the Bloch sphere. For $\mu_1=\nu_+$ and $\mu_1=\nu_-$ the jumping states are generated by $\vec{u}_+$ and $\vec{u}_-$, respectively. Fig.~\ref{2sj}(c)-(d) shows that such solutions lie in the $x=0$ plane. For $\mu_1=\nu_-$, jumping states are spread as far apart as possible; the system spends most of the time in one state, which is nearly aligned with the direction of the steady state, $\vec{r}_{\rm ss}$.  For $\mu_1=\nu_+$, jumping states cluster together around the direction of the steady state, $\vec{r}_{\rm ss}$; the probability of occupying each state is more similar and becomes equal as $\varepsilon$ approaches 0. Since  the probability of occupying each state is no longer equal, the Shannon entropy is less than one. This means that one could store the state of the qubit in \emph{less than} one bit \emph{on average}. That is, one could keep track of the state of a collection of $N$ identically monitored qubits using only $N h$ bits.
Actually, the entropy of the solution due to $\vec{v}_-$  is very low. That is, it is very close to the von Neumann entropy of the steady state, which is a lower bound for the Shannon entropy of any ensemble representing for the steady state. For this solution, the Shannon entropy in a Taylor expansion in terms of $\varepsilon$ is 
\begin{equation} \label{hu-}
h(\vec{u}_-)=(\log_2 10-4 \log_2 \varepsilon) \varepsilon^4-48 (\log_2 \varepsilon)\, \varepsilon^6+O(\varepsilon^8)
\end{equation}
 compared to 
\begin{equation} \label{Sss}
S(\rho_{\rm ss})=(\log_2 10-4 \log_2 \varepsilon) \varepsilon^4+6 (\log_2 \varepsilon)\, \varepsilon^6+O(\varepsilon^8).
\end{equation}

\subsection{Stability of two-state jumping}\label{Sec:Fluorostable2}

For $\mu_1=1/2$, the PR states are $\ket{v_+(\mu_1)}$ and $\ket{v_+(-\mu_1)}$ and the full set of eigenvalues from \erf{eigmu}  can be simplified to 
\begin{equation}
 \lambda_+\!\!\left(\pm\frac{1}{2}\right)=\frac{1}{8}\pm\frac{i \varepsilon}{2} \text{ and }
\lambda_-\!\!\left(\pm\frac{1}{2}\right)=\frac{5}{8}\mp\frac{i \varepsilon}{2}.
\end{equation}
Thus ${\rm Re}\lambda_1^{{ e}} =  {\rm  Re}\lambda_2^{{e}} = 1/8$, while 
${\rm  Re}\lambda_1^{{ o}} =  {\rm  Re}\lambda_2^{{ o}} = 5/8$. Therefore $C = 1/25$, 
and the system is mean-square stable. Moreover, it is piecewise deterministically stable (that is, the fidelity increases monotonically except perhaps at jumps.)

\begin{figure}
\includegraphics[width=.85\columnwidth, clip=true, trim= 0 0 0 0]{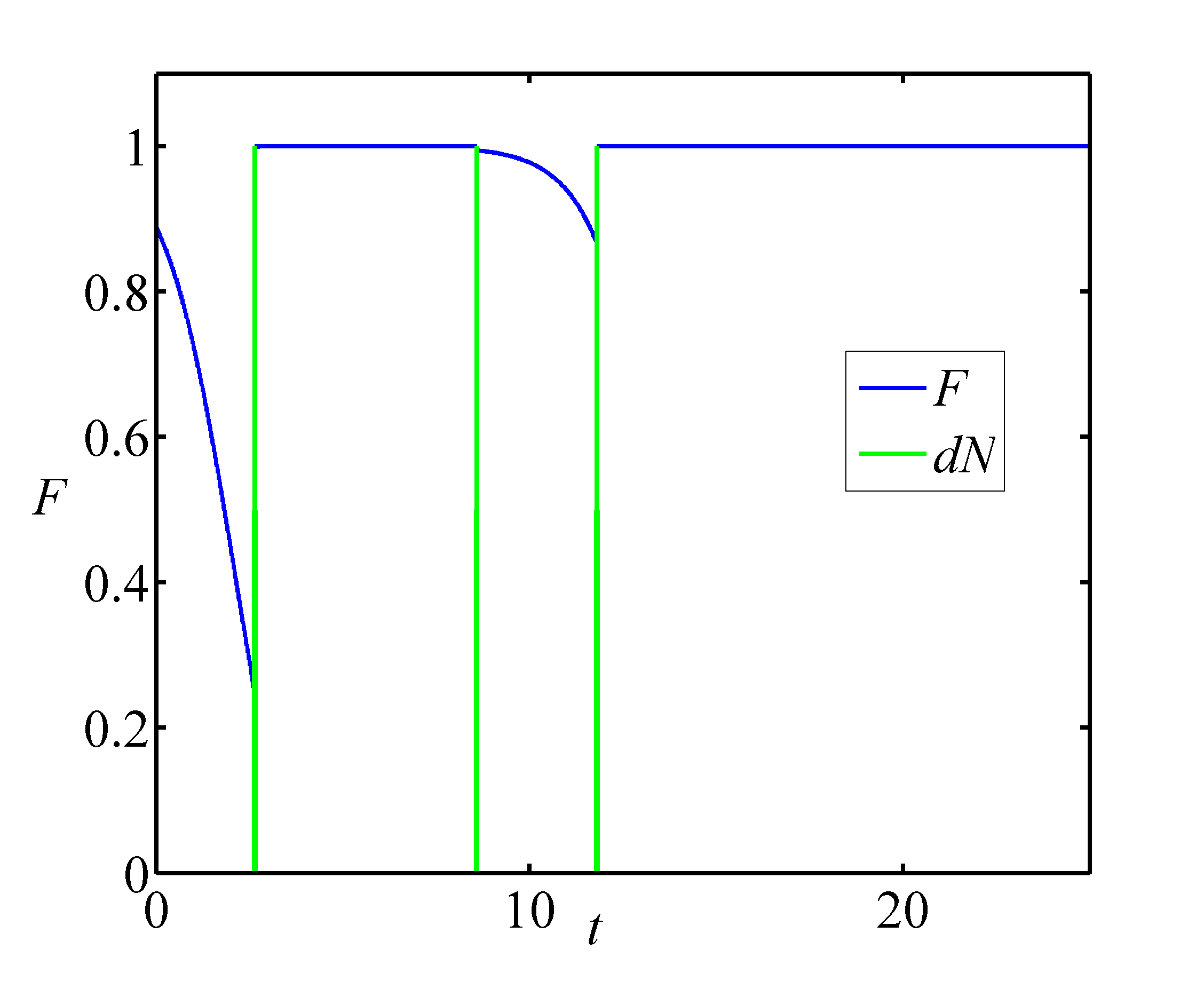} 
\vspace{-2mm}
\caption{\label{unEvol}(Color online.)  Alternating stages of stable and unstable two-jump evolution, exaggerated by an atypical trajectory. The blue (darker)
line represent the fidelity of the state with the ideal PR ensemble and green (lighter) line shows when jumps happen  (photon count increment $dN=1$). This is a highly improbable evolution, as the systems spends  an  uncharacteristically long time in an unstable step and uncharacteristically short time in a stable step. Fidelity is decreasing during the unstable step. At the second jump there is also a decrease in fidelity due to the jump itself. Time $t$ is measured in units of $\gamma^{-1}$.}
\end{figure}

\begin{figure}[htp]
  \begin{center}
    \subfigure{\includegraphics[width=3.75cm, clip=true, trim= 0 0 259 0]{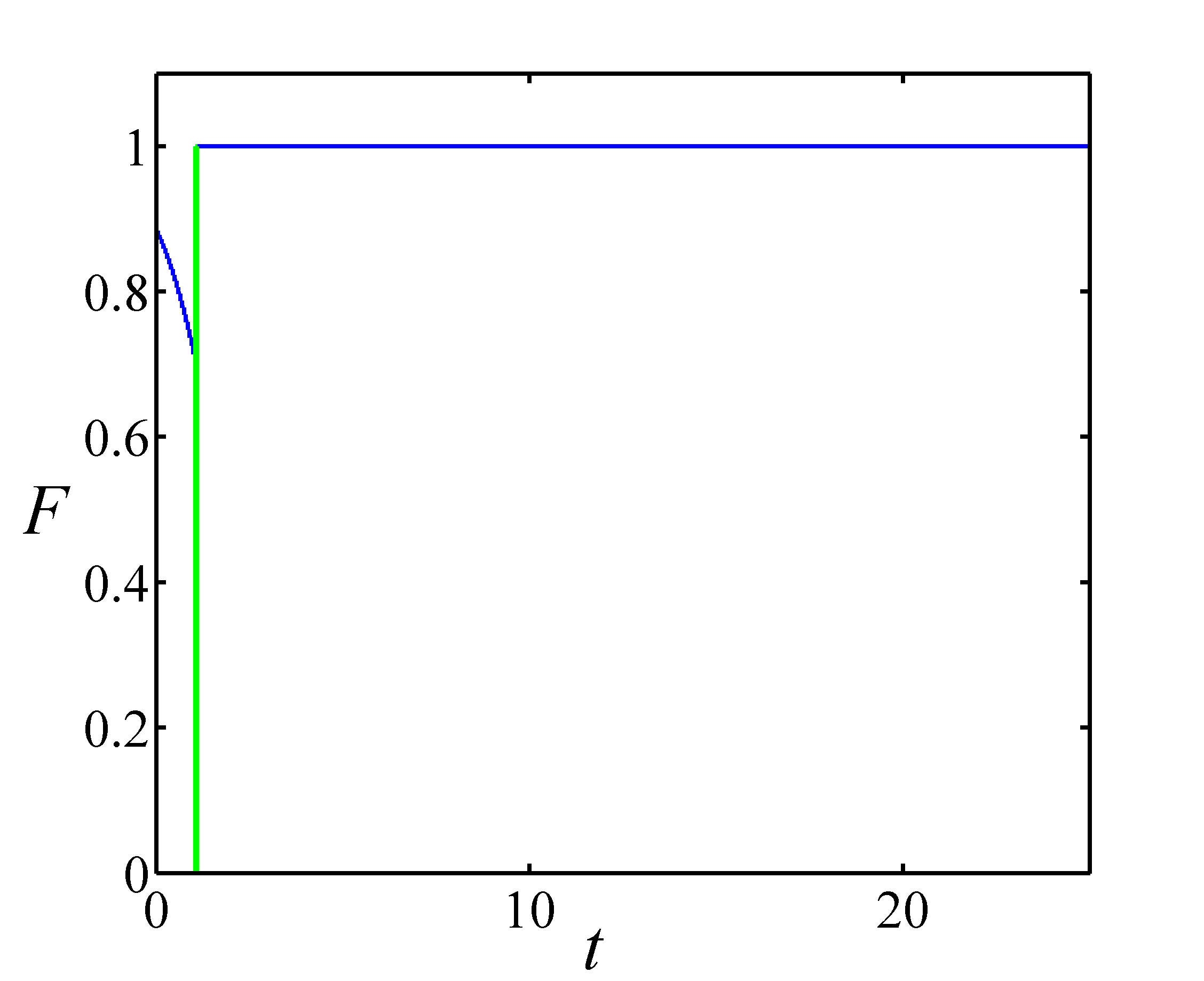}}
$\cdots$
    \subfigure{\includegraphics[width=3.75cm, clip=true, trim= 259 0 0 0]{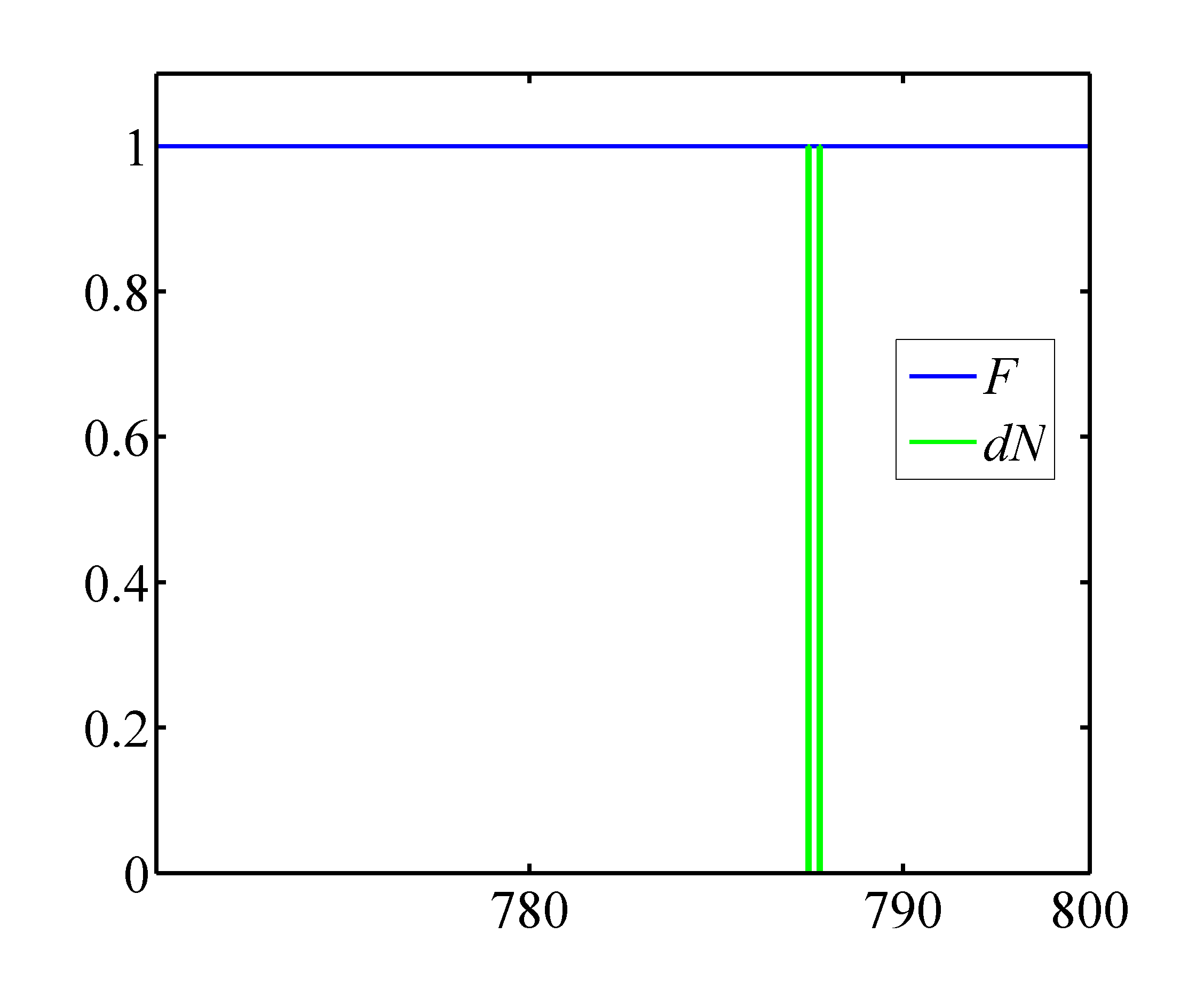}}
  \end{center}
  \caption{(Color online.) As in Fig.~\ref{unEvol}, but showing a typical trajectory. The fidelity decreases visibly only during the first stage, as stable stages last much longer than unstable stages.   }
  \label{typEvol}
\end{figure}

For $\mu_1=\nu_-$, the situation is quite different.  The PR ensemble is comprised from $\ket{v_+(\mu_1)}$ and $\ket{v_-(-\mu_1)}$. It is still mean-square stable, but  the short-lived state $\ket{v_2^e} = \ket{v_-(-\mu_1)}$ has an eigenvalue such that ${\rm Re}\lambda_2^e > {\rm Re}\lambda_2^o$, where $\ket{v_2^o} = \ket{v_+(-\mu_1)}$. That is, every second stage is unstable. We illustrate this situation in figures~\ref{unEvol} and ~\ref{typEvol}. Here we assume that evolution starts with the unstable step; that is, the initial strength of the local oscillator is $-\mu_1$. We also let $\varepsilon=0.1$ and the initial state is $\ket{\tilde{\psi}_0}=-1.09 \ket{v_-(-\mu_1)}+0.5 \ket{v_+(-\mu_1)}$. Figure~\ref{unEvol} shows highly untypical evolution, where the short-lived state $\ket{v_+(-\mu_1)}$ exists for an improbably long time and the long-lived state $\ket{v_-(-\mu_1)}$ lasts an  improbably short time. One can see here that for stages  governed by $-\mu_1$ the fidelity decays as the system moves away from the desired PR ensemble state. 
Figure~\ref{typEvol} shows typical evolution under the same conditions. Here the instability can be seen only during the first step, which is short (at time $\tau_1\approx 1.1 \gamma^{-1}$). The next jump does not happen for long time ( $\tau_2 \approx 792.5\gamma^{-1}$) and it is very short (it lasts for $ \tau_3 \approx  0.3 \gamma^{-1}$). Thus  the unstable stages  contribute very little to the overall evolution of the fidelity. As a result it does not disturb mean-square stability for the system.

\begin{figure}
 \includegraphics[width=.85\columnwidth]{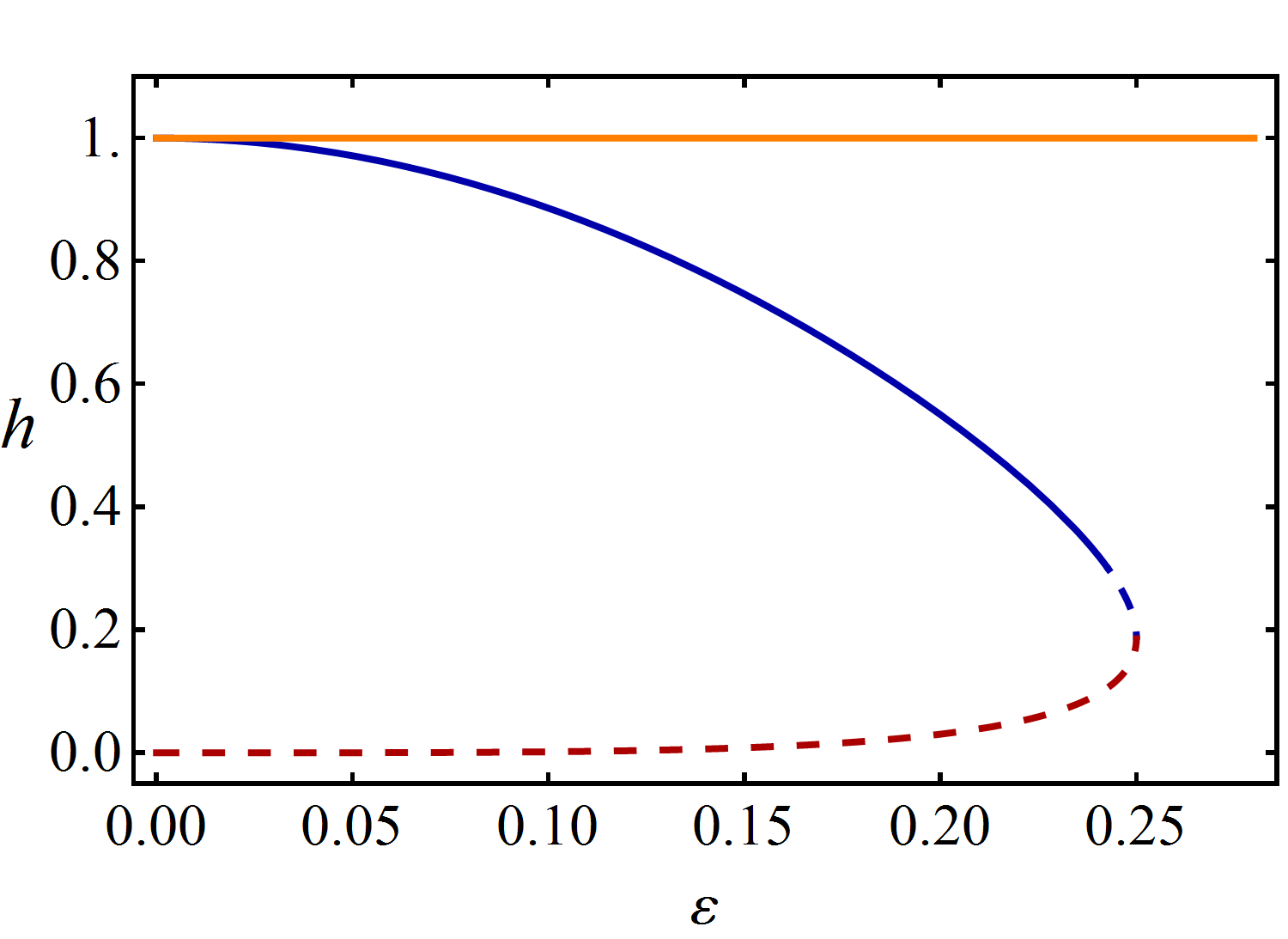} 
\vspace{-2mm}
\caption{\label{entWstab} (Color online.)  Ensemble Shannon entropy $h$ for the three different two-state jumping solutions.   A solid line indicate mean-square stability as well as piecewise deterministic stability. A dashed line indicates that the solution has only mean-square stability, with one of its stages being unstable. The uppermost (yellow) line shows entropy for PR ensemble with $\mu_1=1/2$. The middle line (blue) describe entropy for solutions with $\mu_1=\nu_+$ and the lowest line (red) is for $\mu_1=\nu_-$. }
\end{figure}

For $\mu_1=\nu_+$  the situation is even more curious. For $\varepsilon<\varepsilon_0\approx 0.243$, the PR ensemble is comprised from $\ket{v_+(\mu_1)}$ and $\ket{v_+(-\mu_1)}$ and for $\varepsilon>\varepsilon_0$,  the PR ensemble is generated from $\ket{v_+(\mu_1)}$ and $\ket{v_-(-\mu_1)}$. The system is always mean-square stable, and is piecewise deterministically stable for $\varepsilon<\varepsilon_0$. However for $\varepsilon>\varepsilon_0$ the short-lived stage associated with $\ket{v_-(-\mu_1)}$ becomes unstable. At $\varepsilon=\varepsilon_0$, $\lambda_+(-\mu_1)=\lambda_-(-\mu_1)$. In Fig.~\ref{entWstab}, we show the entropy for three different solutions with two-state jumping and the type of stability they exhibit. 

\begin{figure}
\includegraphics[width=.85\columnwidth]{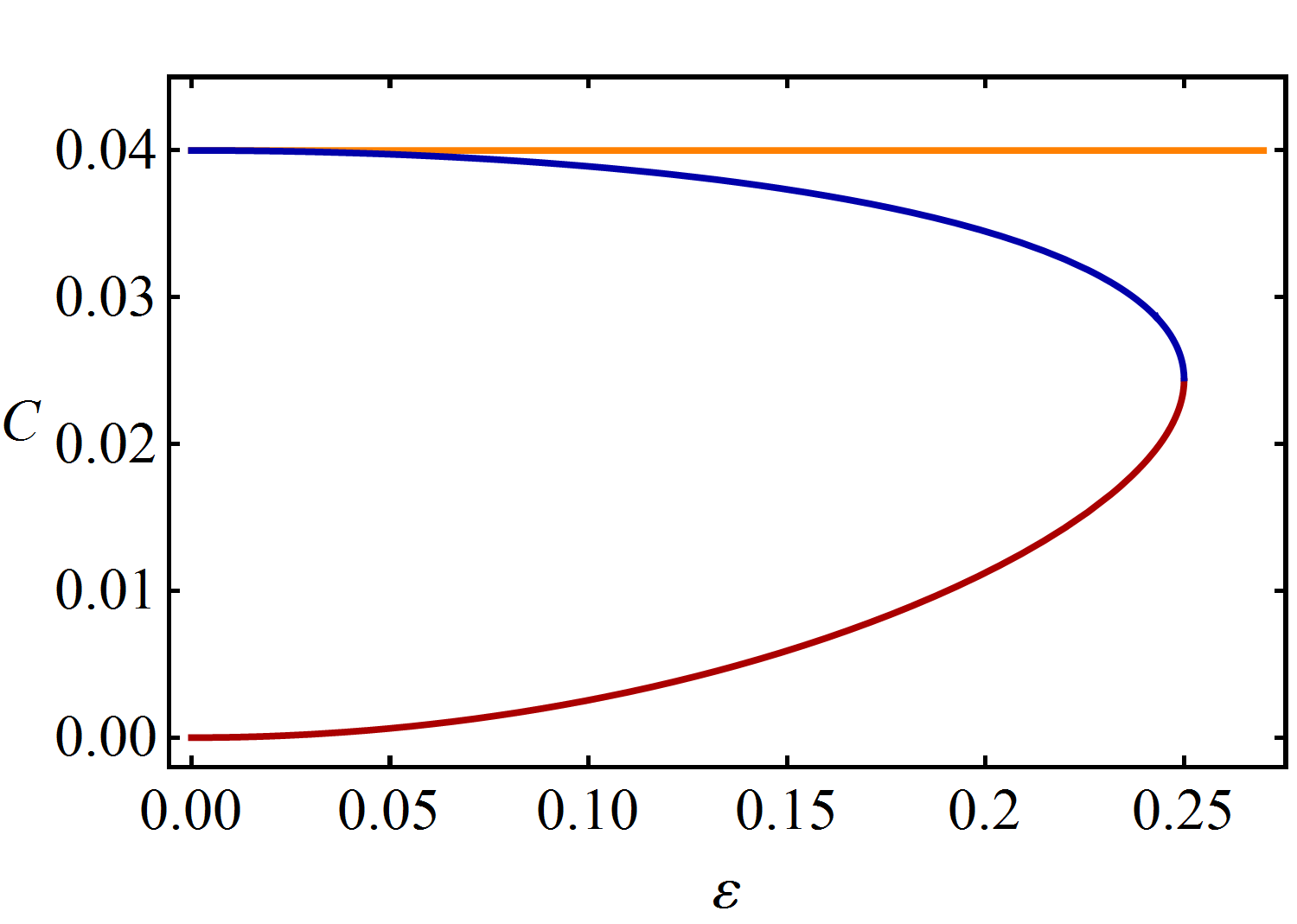} 
\vspace{-2mm}
\caption{\label{convRate}
(Color online.)  The convergence coefficient $C$ as function of $\varepsilon$ for the three different values for $\mu_1$. The yellow (straight) line  corresponds to $\mu_1=1/2$, the red (lowest) line  to  $\mu_1=\nu_-$  and the blue (intermediate) line   to $\mu_1=\nu_+$. Note that the lowest entropy solutions, which is not piecewise deterministically stable, is  the most stable (smallest $C$) in the mean-square sense. }
\end{figure}

Recall from \erf{eq:abeta2l} that the deviation of the fidelity from unity decays as $C^{l}$, 
after $2l$ jumps. The critical constant for all three values of $\mu_1$ can be written as
\begin{equation}
C=\frac{|\mu_1|^4}{4 {\rm  Re}\lambda_1^{{ o}} {\rm  Re}\lambda_2^{{ o}}}.
\end{equation}
We show the convergence coefficient $C$ as a function of $\varepsilon$ for all three possible values of $\mu_1$ in Fig~\ref{convRate}.
There is a clear (although not perfect) correlation between the low entropy solutions and the most quickly converging (in the mean-square sense) solutions. Note however that the most rapidly converging solution is typically {\em not} piecewise deterministically stable.  

We note here that from Eq.~(\ref{eq:msF}), one can see that convergence coefficient $C$ characterizes only certain aspect of mean-square stability. In particular, it determines how many stages are needed for convergence to the PR ensemble. Thus the smaller $C$ in Fig.~\ref{convRate} corresponds to fewer stages needed to achieve certain level of convergence in the mean-square sense. This plot, however, does not tell us about the time needed to achieve convergence. For example, although the solution with $\mu_1=\nu_-$ has a much smaller value for $C$ than $\mu_1=1/2$ solution, and so requires more jumps to converge, it could have shorter stages between every jump so that it approaches fidelity in a shorter time than $\mu_1=\nu_-$ solution. To show that this is a plausible scenario we consider an asymptotic rate of convergence to the PR ensemble, which we define as $R = -\ln(C)/\langle t_\text{as}\rangle$, where $\langle t_\text{as}\rangle$ is the expected value for the duration of the complete cycle of evolution for the corresponding PR ensemble. Because asymptotically the actual ensemble converges to the PR ensemble, and using the law of large numbers, the number of cycles undergone in the limit $t\to\infty$ converges to $t/\langle t_\text{as}\rangle$. Thus from Eq.~(\ref{eq:msF}), in the long-time limit, the fidelity $F(t)$ between the actual conditioned state and the record-determined PR state at time $t$ behaves as 
\begin{equation}
\langle F(t) \rangle \sim 1-|\beta_0|^2 (1-|O_1|^2)\exp(-Rt)
\label{eq:msFt}
\end{equation}

\begin{figure}
\includegraphics[width=.85\columnwidth]{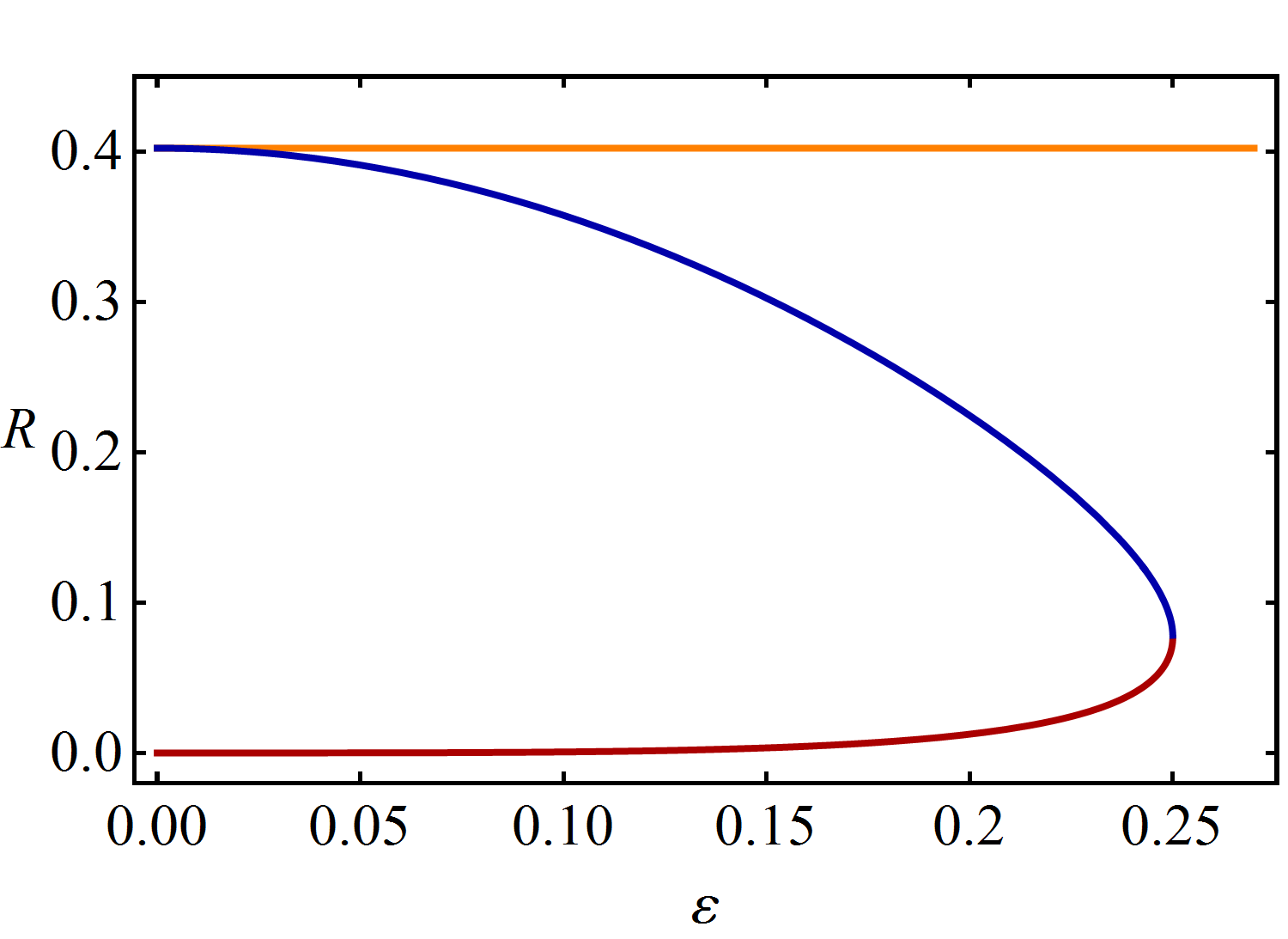} 
\vspace{-2mm}
\caption{\label{TconvRate}
(Color online.)  The asymptotic convergence rate  $R$ as function of $\varepsilon$ for the  three different values for $\mu_1$. The yellow (straight) line  corresponds to $\mu_1=1/2$, the red (lowest) line  to  $\mu_1=\nu_-$  and the blue (intermediate) line   to $\mu_1=\nu_+$.  
Note that, a high value of $R$ indicates faster convergence (per unit time), so this measure of convergence reverses the order of the three schemes relative to that in Fig.~\ref{convRate}.}
\end{figure}

For two-state jumping this asymptotic convergence rate is 
\begin{equation}
R = -\ln(C)/\left[(2{\rm Re \lambda^e_1})^{-1}+(2{\rm Re \lambda^e_2})^{-1}\right],
\end{equation}
and we plot it in Fig.~\ref{TconvRate} as a function of $\varepsilon$ for all three values of $\mu_1$. Here the solution with $\mu_1=1/2$ has the highest asymptotic convergence rate,  with $R=\ln(5)/4$,   
while the solution $\mu_1=\nu_-$ has the lowest, with $R \to 0$ as $\varepsilon \to 0$. 
Specifically, for $\mu_1=\nu_-$, although $\log(C) =O(\log(\varepsilon^2)) \gg 1$, ${\rm Re}\lambda^e_1 = O(\varepsilon^4)$ so that $R = O(\varepsilon^4\log(\varepsilon^2)) \to 0$. This is a complete reversal from the results reported in Fig.~\ref{convRate},  based on $C$. 

The vast difference in rates in the limit $\varepsilon \to 0$ can be understood from the nature of the ensembles as illustrated in Fig.~\ref{2sj}. In all cases the  the average $\rho_{\rm ss}$ differs from $|0\rangle\langle0|$ only at  $O(\varepsilon)$,  and 
differs from a pure state only at   $O(\varepsilon^4)$,  as reflected in \erf{Sss}. In the case $\mu=1/2$, the  ensemble comprises  a pair of Bloch vectors located near $\vec{r}_{\rm ss}$ in a symmetric fashion in the plane perpendicular to $x=0$ plane. For $\mu_1=\nu_+$  the Bloch vectors lie in the $x=0$ plane around $\vec{r}_{\rm ss}$ in a nearly symmetric fashion. In both cases, the system spends the same time in each state and thus the expected time for the duration of each cycle is independent of $\varepsilon$, so the rate of convergence is non-zero. For $\mu_1=\nu_-$, one of the jumping states is nearly aligned with $\rho_{\rm ss}$ [i.e. it has the form  $|0\rangle+ O(\varepsilon)\ket{1}$] and the other state approaches $|1\rangle$ as $\varepsilon\rightarrow0$. For small $\varepsilon$, the system  spends almost all  of the time in the first state. It jumps to the excited state with a rate $O(\varepsilon^4)$, and jumps back with a finite rate. The former process is the rate-limiting step, so the system has a cycle  whose duration tends to infinity as $\varepsilon\rightarrow0$,  giving a  convergence rate near $0$. 


However, we need to be careful in interpreting the results from Fig.~\ref{TconvRate} as it only shows the  asymptotic rate.
Most of the convergence happens in the initial evolution --- during the first few stages, while the state of the system is drastically different from the PR ensemble.  The asymptotic rate becomes relevant only when the state of the system closely resembles  the  PR ensemble and, in a way, the convergence has already occurred. We, therefore, also consider another measure for the rate of convergence: $R_1 = -\ln(C)/\langle T_1\rangle$. Here $\langle T_1\rangle$ is the expected time for the duration of the first cycle. For two-state jumping, this quantity was investigated numerically.  

Our numerics reveal that the initial convergence rate $R_1$ shows strong dependence on the initial condition. 
For some initial conditions, $R_1$ for the $\mu_1=\nu_+$ solution is above that for the $\mu_1=1/2$ 
solution for all $\varepsilon \leq 1/4$. Because the $\nu_-$ and $\nu_+$ solutions coincide for $\varepsilon = 1/4$, this implies that, for these initial conditions, and for $\varepsilon$ sufficiently close to $1/4$, the $\mu_1=1/2$ solution has the slowest initial rate of convergence for some range of $\varepsilon$  near $1/4$.  On the other hand, for other initial conditions, $R_1$ for the $\mu_1=\nu_+$ solution is below that for the $\mu_1=1/2$ solution for all $\epsilon \leq 1/4$, and the same ordering as seen in the asymptotic limit (Fig~\ref{TconvRate}) occurs for some range of $\varepsilon$ near $1/4$. Because $-\ln(C)$ diverges (slowly) to $+\infty$ as $\varepsilon \to 0$ for $\mu_1=\nu_-$, for a fixed initial condition this last solution always  has the largest initial rate of convergence  $R_1$ as $\varepsilon \to 0$. 

\begin{figure}
\includegraphics[width=.85\columnwidth]{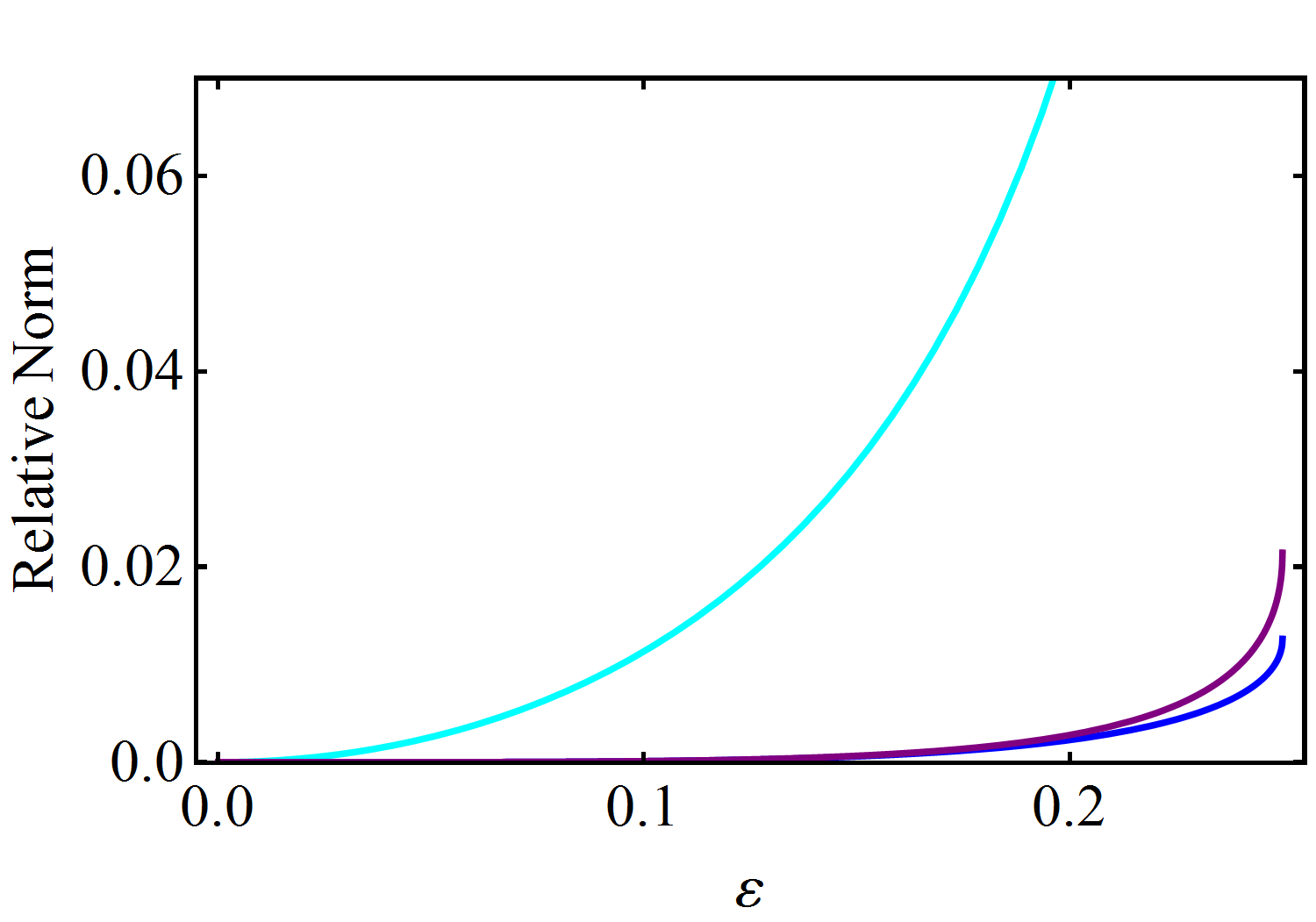} 
\vspace{-2mm}
\caption{\label{FidafterJ}
(Color online.)  Quantities relevant to a fidelity decrease upon jumping in the two-state jumping solution with $\mu_1=\nu_-$. The blue (lowest) line shows the achievable lower bound for the relative norm  $||\ket{\tilde{\phi}_2}||^2 / ||\ket{\tilde{\phi}_1}||^2$. The cyan (uppermost) line shows the bound $B$ the jump generated by $\hat{s}_1$ and purple (middle) line is the same bound for jumps generated by $\hat{s}_2$. Since the curves for $B$ are both above the lower bound, it is possible for the system to experience a decrease in fidelity under either type of jump.}
\end{figure}

The last question we address in this section is the effect jumps have on fidelity. For the model we are considering  we can easily evaluate  the upper and lower bounds in Eq.~(\ref{eq:lubound}) to be
\begin{align}
 \lambda_{\rm  min}&=|\mu_1|^2+1/2-\sqrt{|\mu_1|^2+1/4}\\
\lambda_{\rm  max}&=|\mu_1|^2+1/2+\sqrt{|\mu_1|^2+1/4}
\end{align}
For $\mu_1=1/2$, these give the bounds 
\begin{equation}
0.043\approx\frac{3}{4}-\frac{1}{\sqrt{2}}\leq\frac{||\ket{\tilde{\phi}_2}||^2}{||\ket{\tilde{\phi}_1}||^2}<\frac{3}{4}+\frac{1}{\sqrt{2}}\approx 1.46.
\end{equation}  
To calculate $B$  from Eq.~(\ref{eq:fjcond}), we observe that for $\mu_1=1/2$ we have 
observe that  $|O_1|=|O_2|$, $|\beta_2|=|\mu_1|\times|\beta_1|$. This gives 
$B=|\beta_2|^2/|\beta_1|^2=1/4$. Because this is greater than $0.043$ it follows that 
sometimes jumps can decrease the fidelity.
For  $\mu_1=\nu_-$, we determine $B$ and the bounds numerically. We find similar results, as plotted in Fig.~\ref{FidafterJ}. This time however $B$ depends on which jump operator one is considering, and a fidelity-decreasing jump is much less likely following a stage of {\em unstable} evolution.  Our results confirms what was seen in the second jump in the untypical evolution depicted in Fig.~\ref{unEvol}, where a jump following a stable stage of evolution visibly decreased the fidelity. 

\subsection{Three-state jumping}
\label{3stateJump}
We now consider three state PR ensembles with cyclic jumps as in Ref.~\cite{KarWis11}, still for a qubit subject to resonance fluorescence. In this case, in Eq.~(\ref{bvJumpCond})  only $\kappa_{12}$, $\kappa_{23}$ and $\kappa_{31}$ are nonzero and  \erfs{bvNormCond}{bvJumpCond} yield a total of $12$ equations ($9$ equations for jumping conditions and $3$ equations for normalization condition) and $12$ unknowns. Since this system involves quadratic equations,  simple analytic solutions no longer exist.
We find \emph{all} three-state cycles by numerical search for all real solutions to \erfs{bvNormCond}{bvJumpCond} using symbolic-numerical algorithms based on computing a  Groebner basis \cite{CLO,  Cox}, as described in detail in Appendix~\ref{app:GroebnerBasis}. 

One of the reasons to study $K$-state PR ensembles for $K>2$  is the search for low entropy solutions that allow for efficient tracking.  As we have shown, for $\varepsilon>1/4$ there are no low entropy solutions for $K=2$.  
However by moving to $K=3$ we open the possibility for ensembles with different properties. 
The intuitive reason for the greater flexibility is as follows. Recall that for two-state jumping, we could use only  one eigenvector of $A$ to generate  the PR states  and we could not use complex eigenvectors because Bloch vectors must have real components. Observe that complex eigenvectors of $A$ come in conjugate pairs. Thus, for three-state jumping, we can construct (real) Bloch vectors for PR states if we use conjugate pairs of eigenvectors of $A$. This method imposes additional constraints  however and cannot be used for all possible $\varepsilon$. We now explore when complex eigenvectors can yield a PR ensemble.

We first  show that any cyclic jumps between three states is generated from two eigenvectors of $A$. We observe that Eqs.~(\ref{bvJumpCond}) imply that for $\vec{s}_j=\vec{r}_j-\vec{r}_{\rm ss}$,  with $j\in \{1,2,3\}$,  
\begin{equation}
(A-\kappa_{12})(A-\kappa_{23})(A-\kappa_{31})\vec{s}_j =-\kappa_{12}\kappa_{23}\kappa_{31}\vec{s}_j
\end{equation}
Since $A$ is invertible, we can reformulate this condition as $g(A)\vec{s}_j=0$, where we have defined a quadratic function $g(A)=A^2 - (\kappa_{12} + \kappa_{23} + \kappa_{31})A + (\kappa_{12} \kappa_{23} +\kappa_{12} \kappa_{13}+\kappa_{23} \kappa_{31})$. 
This is an eigenvalue equation for $\vec{s}_j$. Note that all eigenvectors of $A$ are also eigenvectors of $g(A)$ and vice verse. Thus $\vec{s}_j$ is a linear combination of eigenvectors of $A$, whose eigenvalues $\lambda$ satisfy equation $g(\lambda)=0$. If all eigenvalues of $A$ are distinct, then only two eigenvalues of $A$ can satisfy equation $g(\lambda)=0$ and only two eigenvectors of $A$ are used to construct $\vec{s}_j$. 

However, not every conjugate pair of complex eigenvectors of $A$ can be used to construct $\vec{s}_j$. Suppose $A$ has two complex eigenvalues $\lambda$ and ${\lambda}^*$. Equation $g(\lambda)=0$ implies that $\kappa_{12} + \kappa_{23} + \kappa_{31}=2 {\rm Re}\lambda$ and $\kappa_{12} \kappa_{23} +\kappa_{12} \kappa_{13}+\kappa_{23} \kappa_{31}=|\lambda|^2$.  By construction, $\kappa_{12},\, \kappa_{23},\,\kappa_{31}$ are real, which is only possible if $({\rm Re}\lambda)^2>3({\rm Im}\lambda)^2$. In particular, for the resonance fluorescence example, complex eigenvectors cannot be used to generate three-state PR states  if $|\varepsilon|>1/2$. This  bound is actually not tight as was shown by numerical search for solutions to \erfs{bvNormCond}{bvJumpCond} for three-state jumping \cite{KarWis11}. These determined that such solutions exist iff $|\varepsilon|\leq 0.282$. 

\begin{figure}
\subfigure[]{
 \includegraphics[width=3.75cm, 
trim=10 10 10 10]{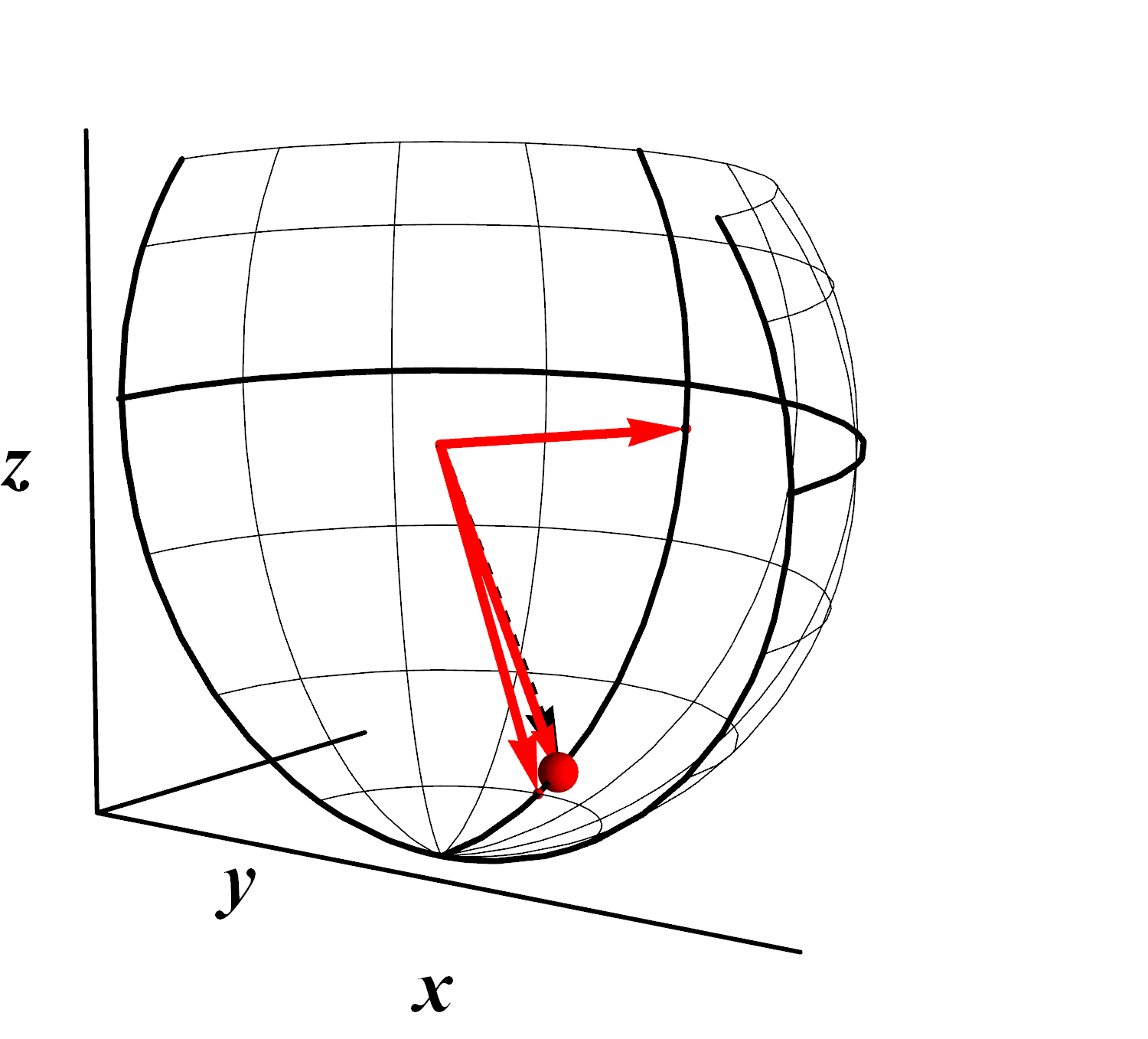}
}
\subfigure[]{
 \includegraphics[width=3.75cm, 
trim=10 10 10 10]{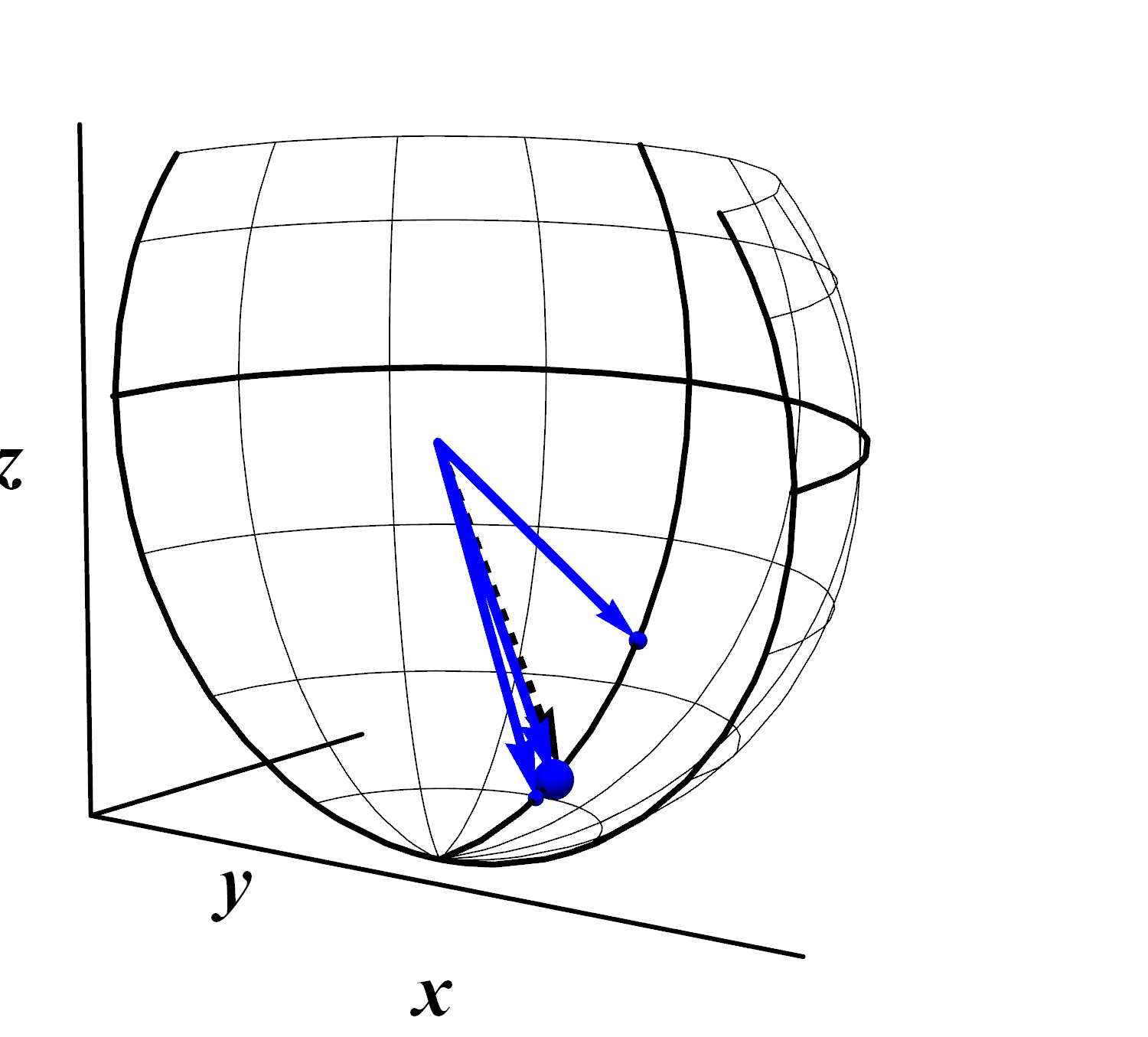}}
\caption{ (Color online.) Solid arrows show Bloch vectors for 3-state jumping. The volume of the sphere at the tip of each arrow represents the probability that the qubit occupies the corresponding pure state.  The dashed arrow is $\vec{r}_{\rm ss}$. Solutions in (a) and (b) are shown for $\varepsilon=0.27$. These are low entropy solutions for 3-state jumping that are generated from complex conjugate pair of eigenvectors of $A$. Note that low entropy solution for 2-state jumping does not exist for such $\varepsilon$.  The solution in (a) has lower entropy than in (b).  }
\label{p27}
\end{figure}

\begin{figure}
\subfigure[]{
 \includegraphics[width=3.75cm, 
trim=10 10 10 10]{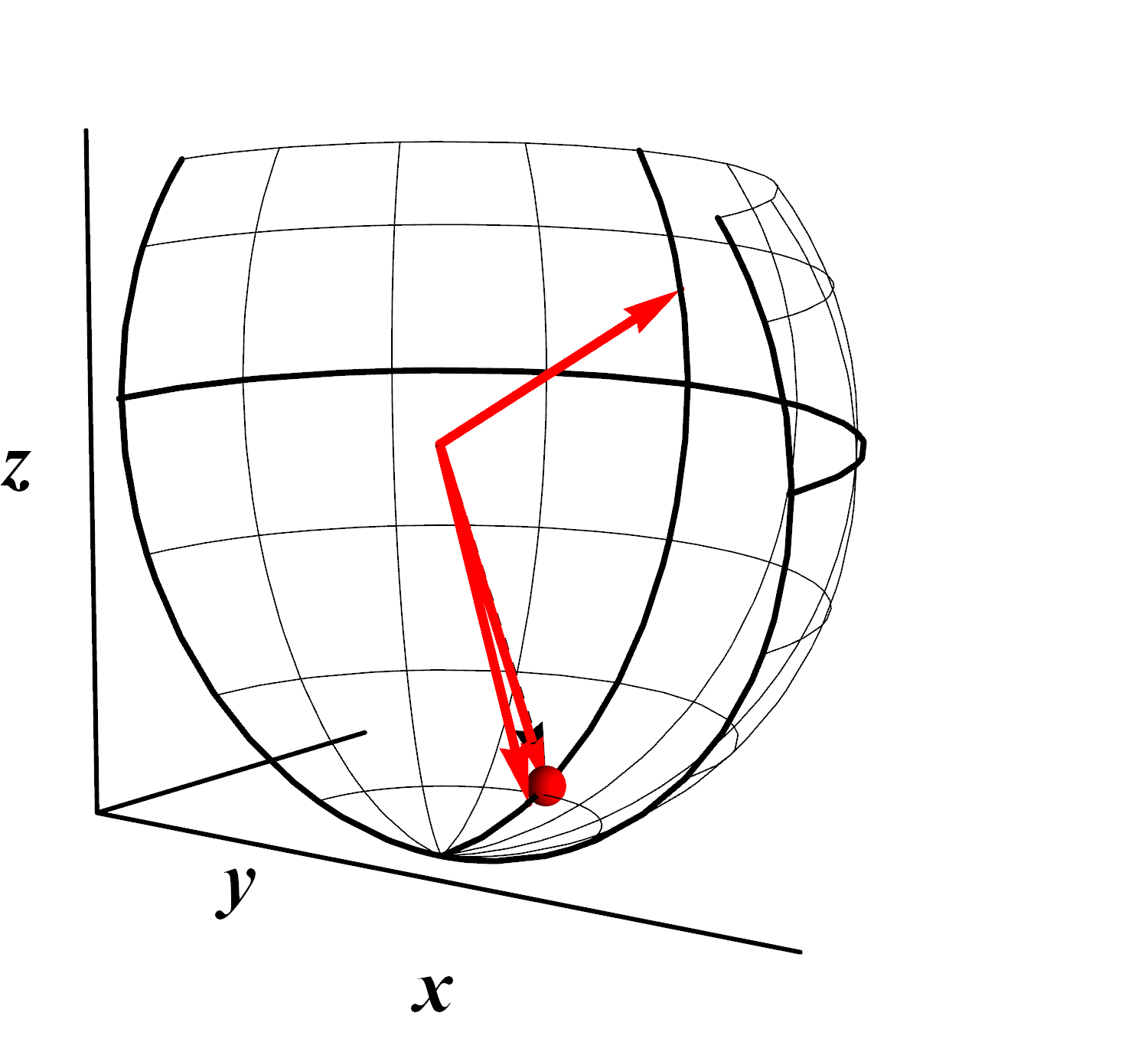}
}\hspace{-2.5 mm}%
\subfigure[]{
 \includegraphics[width=3.75cm, 
trim=10 10 10 10]{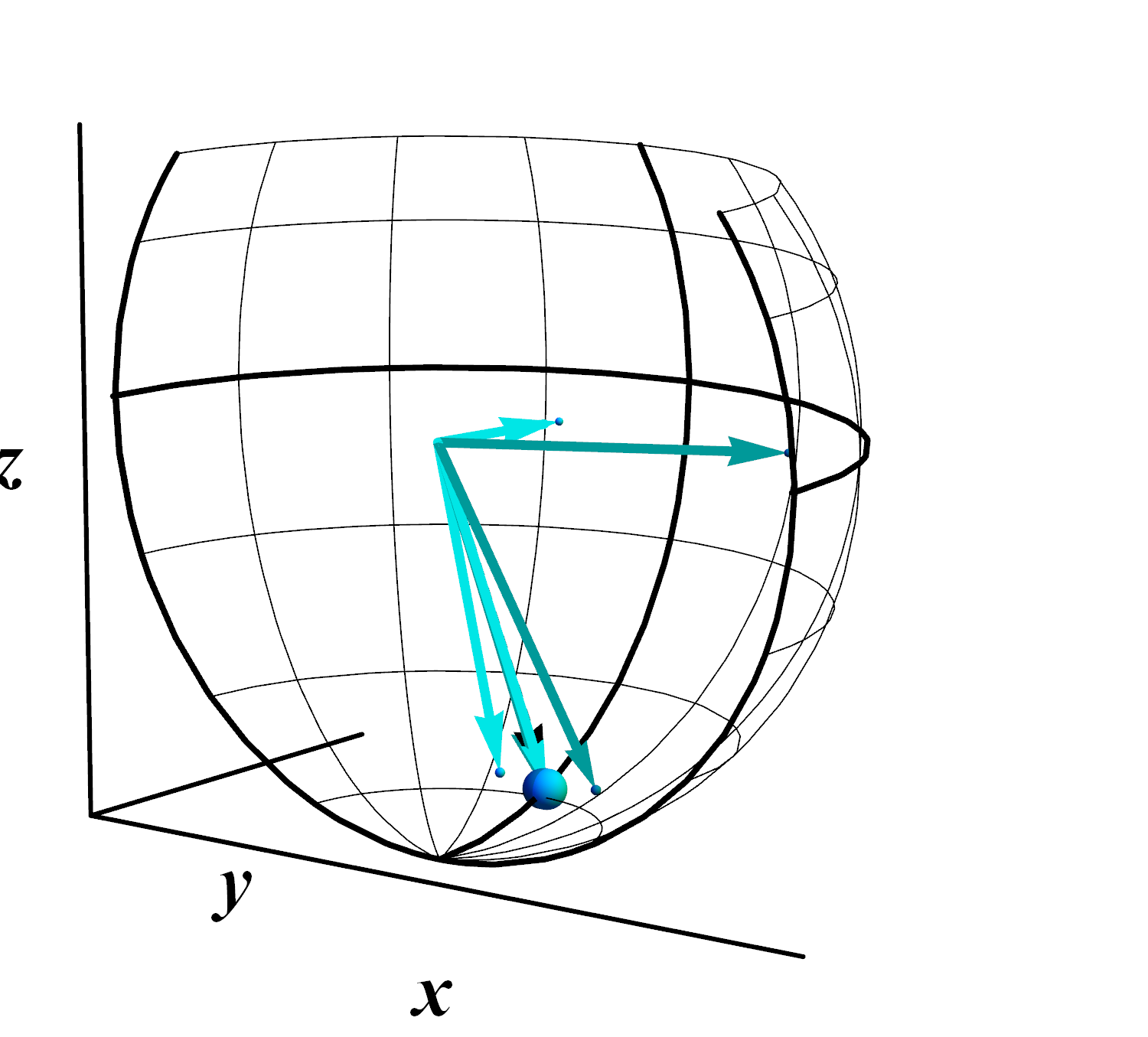}
}\vspace{-2 mm}
\subfigure[]{
 \includegraphics[width=3.75cm, 
trim=10 10 10 10]{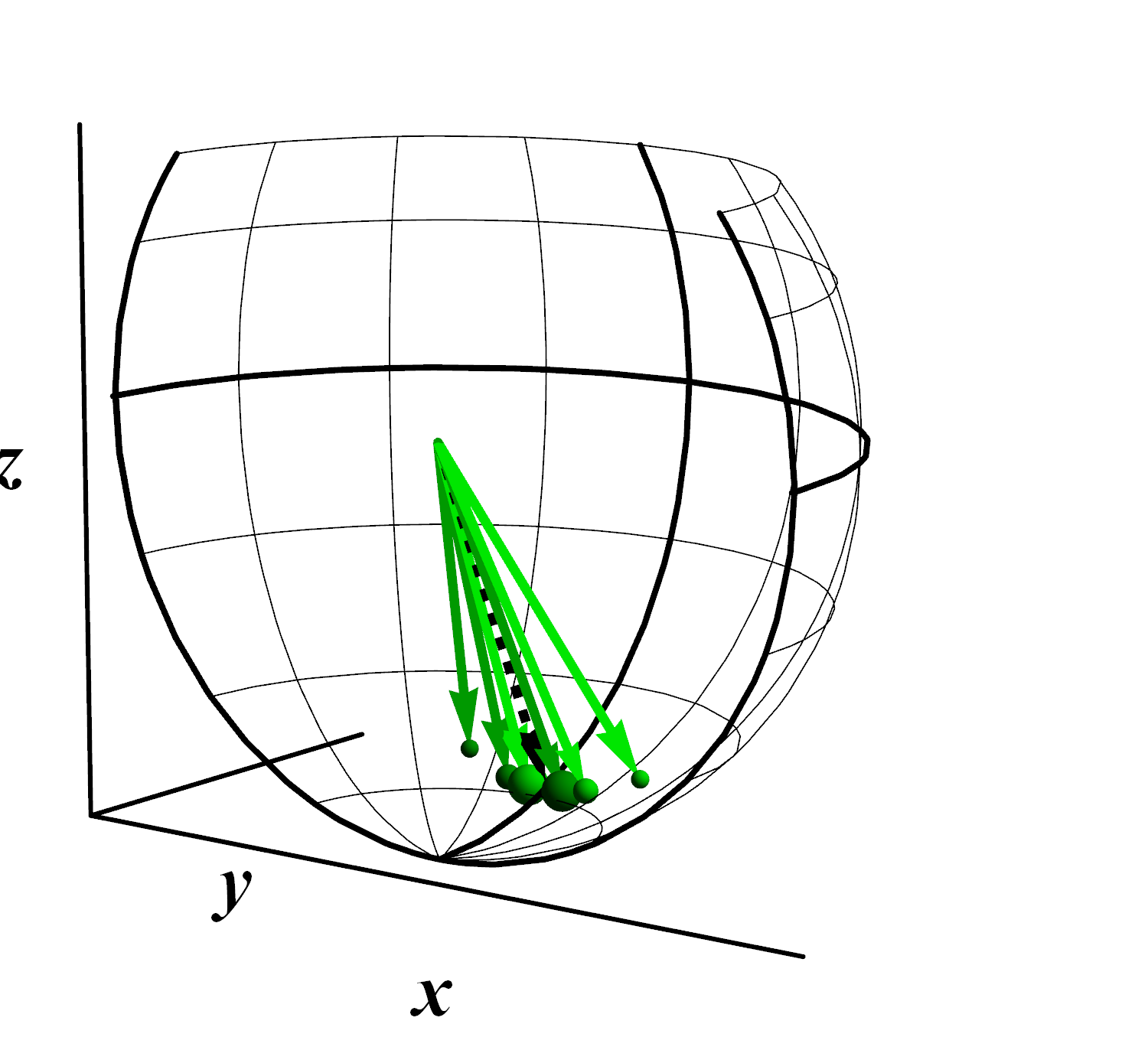}
}\hspace{-2.5 mm}%
\subfigure[]{
 \includegraphics[width=3.75cm, 
trim=10 10 10 10]{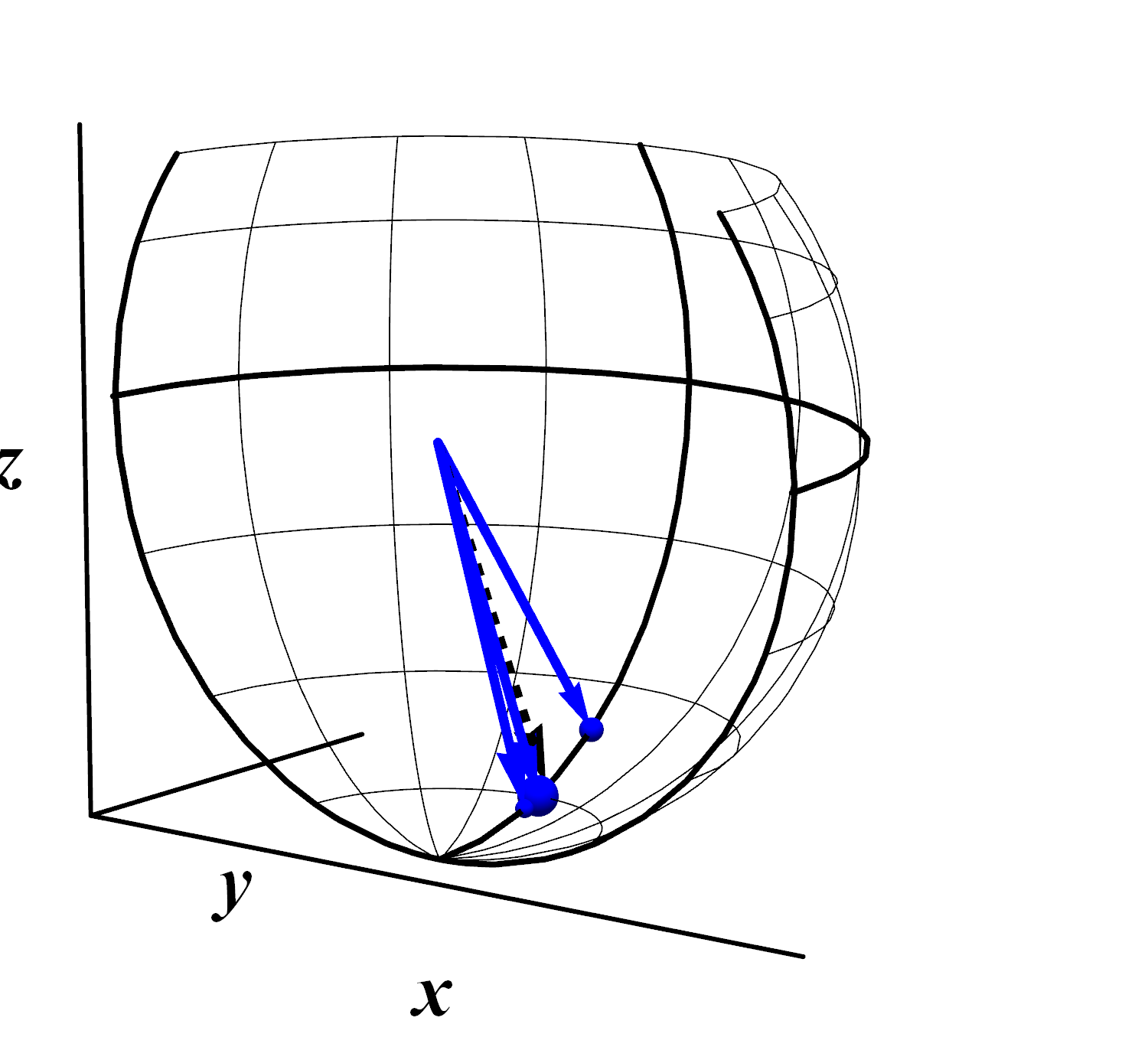}
}\vspace{-5mm}
\caption{(Color online.) As in Fig.~\ref{p27}, but showing all 3-state jumping solutions for $\varepsilon=0.23$. The solution in (a) has the smallest entropy. Every consequent solution has a larger entropy. Solutions (a) and (d) correspond to (a) and (b) in Fig.~\ref{p27}. Solutions (b) and (c) are actually pairs of solutions, symmetric about the $x=0$ plane.}
\label{p23}
\vspace{-5mm}
\end{figure}

\begin{figure}
\centering
\subfigure[]{
 \includegraphics[width=3.7cm, 
trim=10 10 10 10]{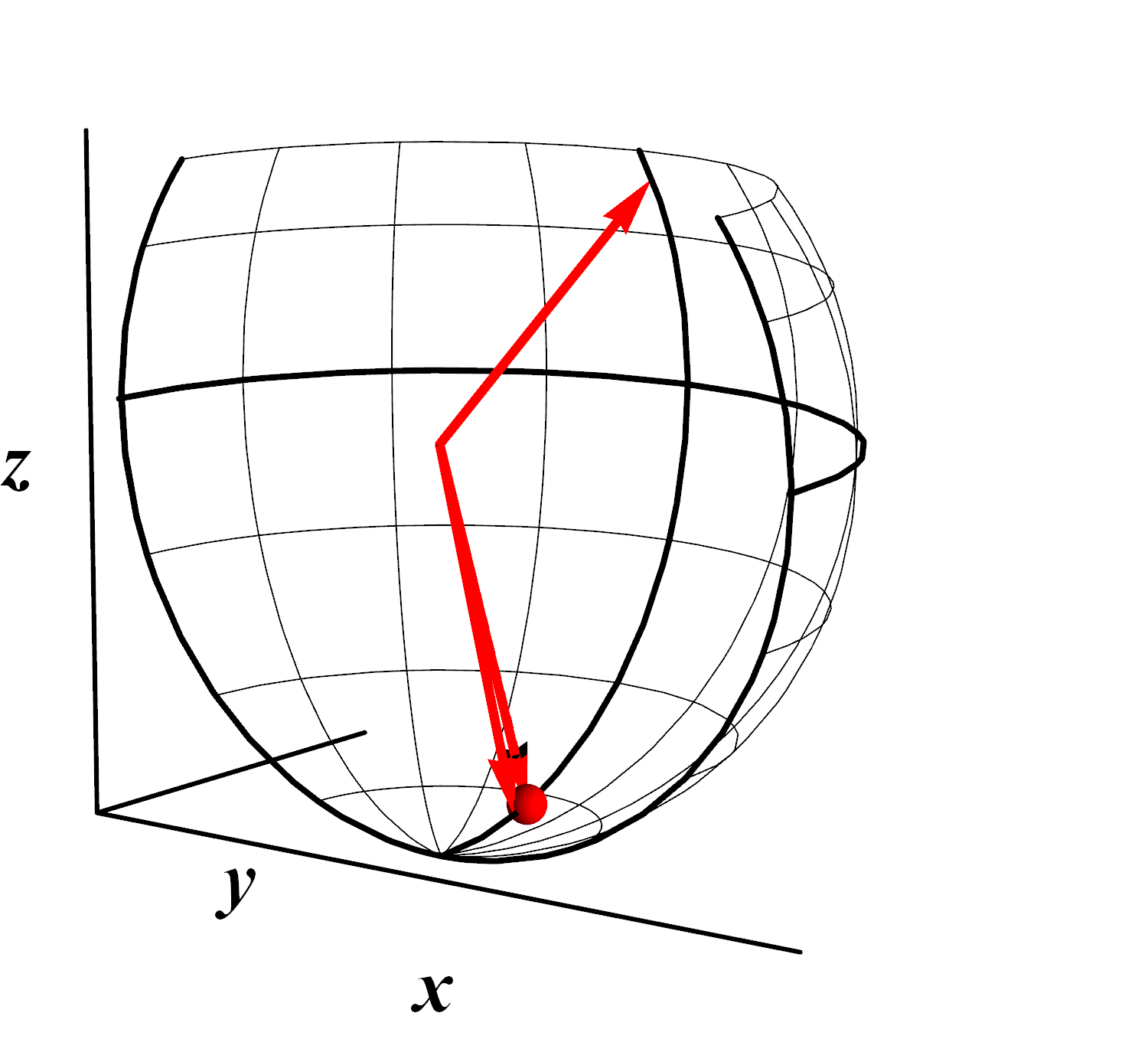}
} 
\subfigure[]{
 \includegraphics[width=3.7cm, 
trim=10 10 10 10]{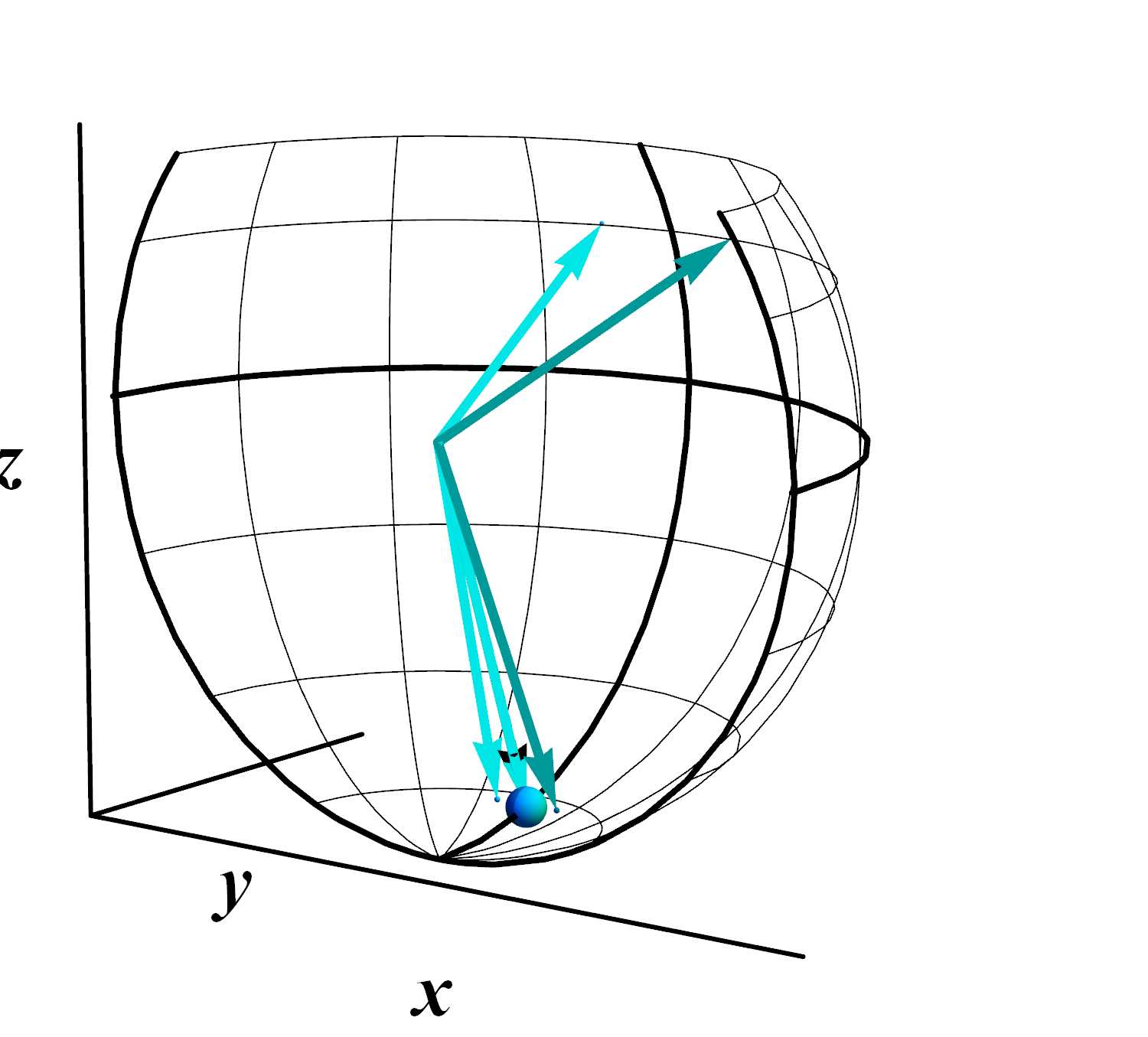}
}
\subfigure[]{
 \includegraphics[width=3.7cm, 
trim=10 10 10 10]{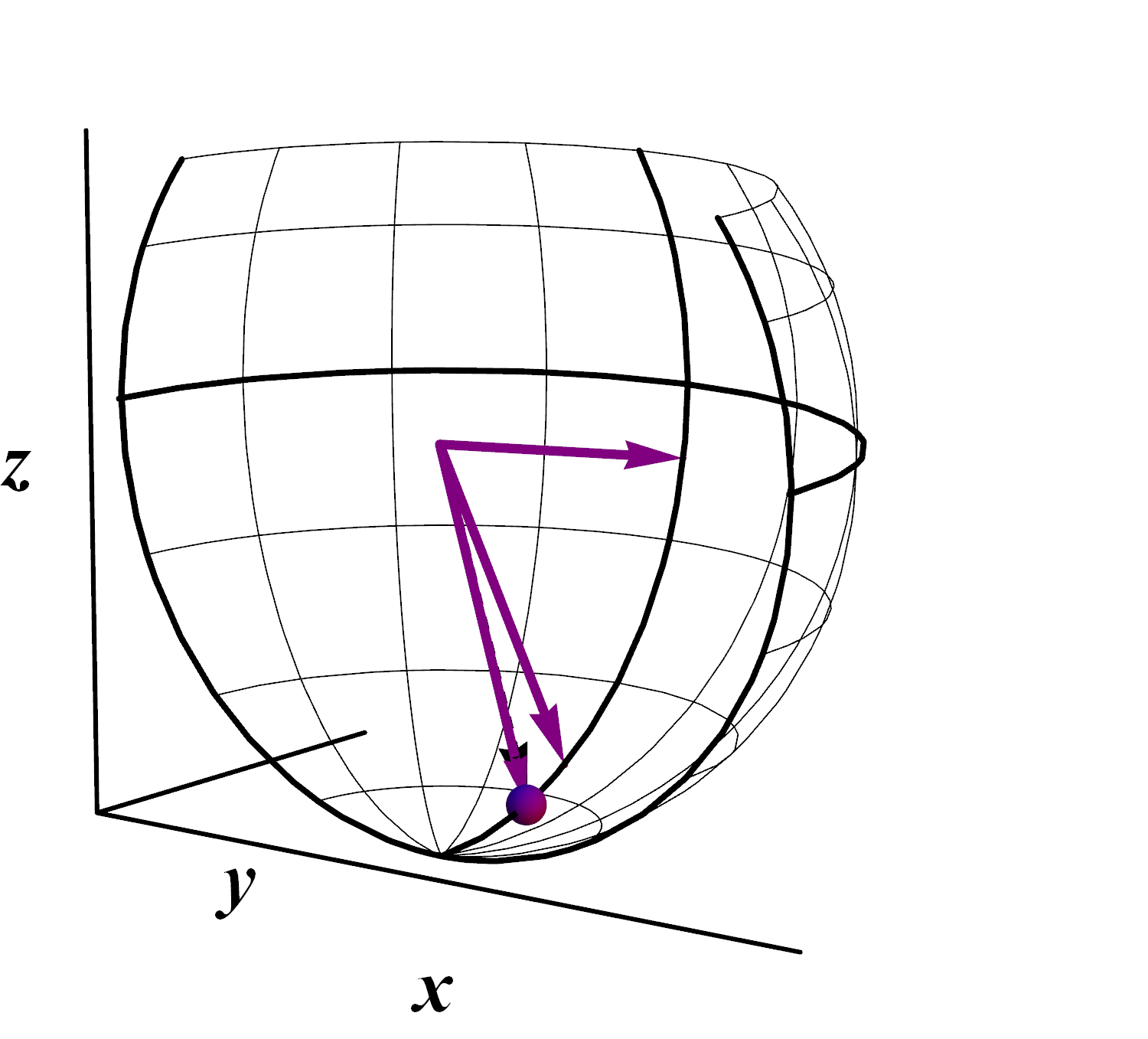}
}
\subfigure[]{
 \includegraphics[width=3.7cm, 
trim=10 10 10 10]{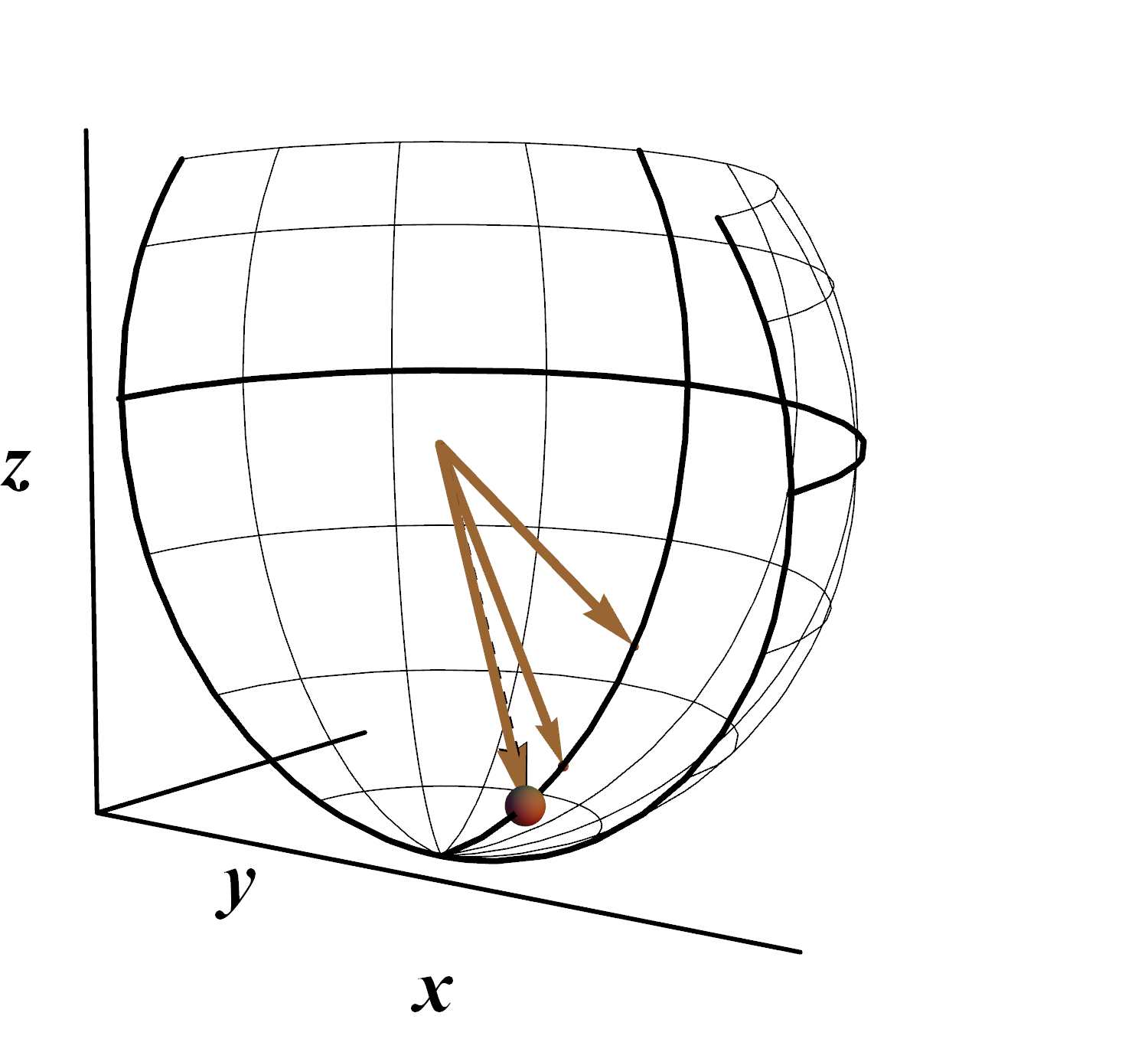}
}
\subfigure[]{
 \includegraphics[width=3.7cm, 
trim=10 10 10 10]{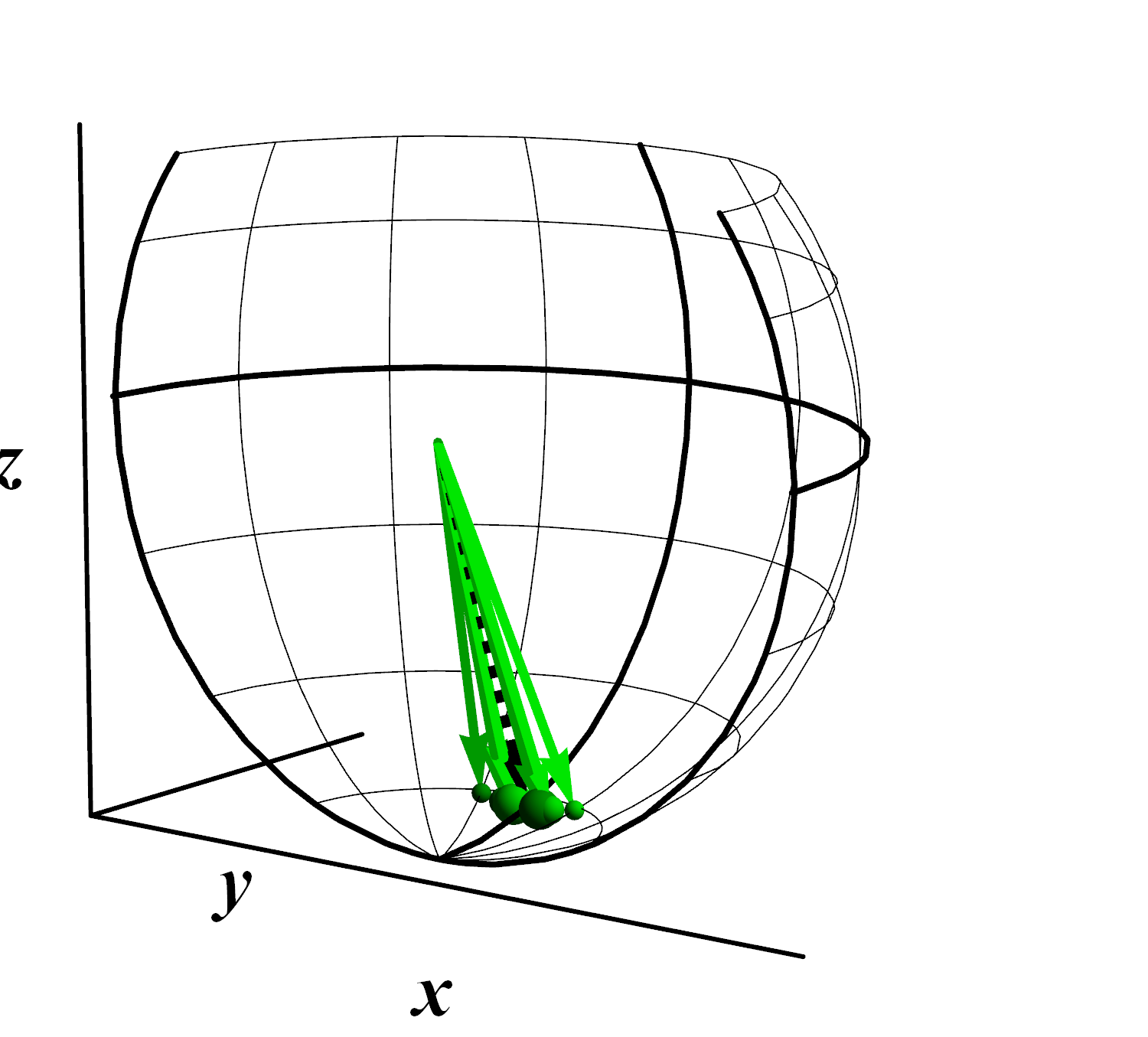}
}
\subfigure[]{
 \includegraphics[width=3.7cm, 
trim=10 10 10 10]{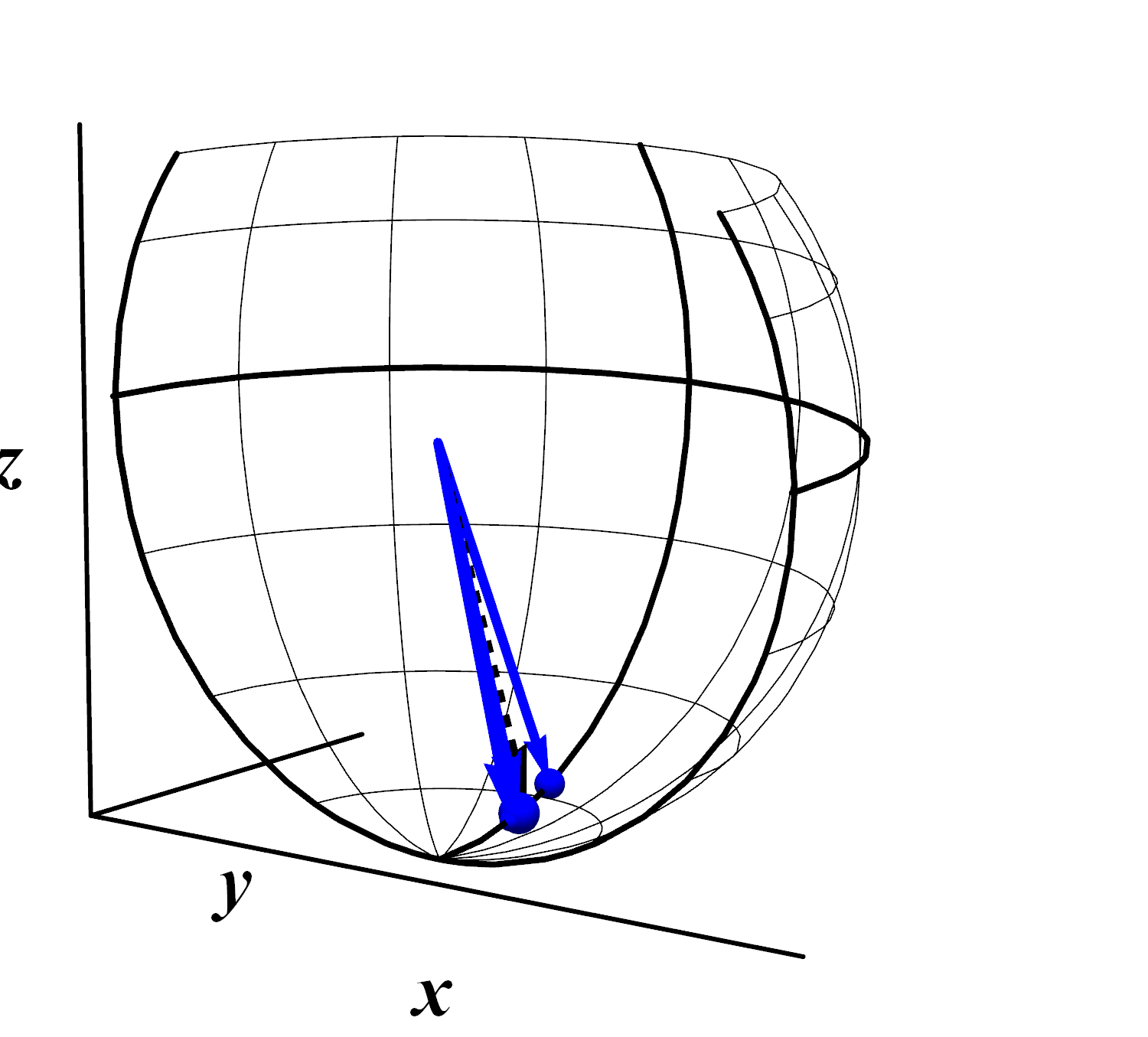}
}
\caption{(Color online.) As in Fig.~\ref{p23}, but showing all 3-state jumping solutions for $\varepsilon=0.18$. Again, the solution in (a) has the smallest entropy and entropy increases with every consequent solution. Solutions (a) and (b) correspond to (a) and (b) in Fig.~\ref{p23}. Solutions (e) and (f) correspond to (c) and (d) in Fig.~\ref{p23}.}
\label{p18}
\end{figure}

Solutions for cyclic three-state jumping come in pairs. As $|\varepsilon|$ approaches $0$, the Shannon entropy $h$  
for half of the solutions approaches 1.2, whereas $h$ for the other half approaches 0. 
For $0.247<|\varepsilon|\leq 0.282$, there are two preferred ensembles generated from complex eigenvectors $\vec{u}_\pm$ of $A$ and shown in  Fig.~\ref{p27} with $\varepsilon=0.27$.  In the region  $0.183<|\varepsilon|\leq 0.247 $, there are 6 solutions, which are shown in Fig.~\ref{p23} with $\varepsilon=0.23$. Solutions with the same entropy are shown in  one subplot. These entropy-degenerate solutions are mirror images with respect to $x=0$ plane and are constructed from $\vec{u}_1$ and $\vec{u}_-$. The other two solutions have lowest and highest entropy and are generated  by $\vec{u}_+$ and $\vec{u}_-$.   
In the last region, $|\varepsilon|<0.183$, there are 8 solutions, which are shown in Fig.~\ref{p18} with $\varepsilon=0.18$. Solutions with the same entropy still appear in the same subplot and are mirror images with respect to $x=0$ plane. All other solutions are constructed from $\vec{u}_+$ and $\vec{u}_-$.  
Solutions that have unique entropy always lie in $x=0$ plane.

In ensemble $\{\wp_k,\,\vec{r}_k\}_{k=1,2, 3}$, we can order the states by probability $\wp_k$ in decreasing order. The system always jumps from one state to another in this order. 
For solutions with lower entropy, the systems spends most of the time in state one, which is nearly aligned with the steady state. For the lowest entropy ensemble, the  states are spread out on the Bloch sphere as far away from each other as possible. As the entropy increases, states with small probability tend to move closer to the steady state. And for high entropy ensembles, the probability for occupying each state tends to equalize, and  none of the states align with the steady state, but instead  cluster around it. In Table~\ref{table1}, for $\varepsilon=0.15$ we report the geometric characteristics of the six distinct solutions. 
The total angle $\angle S_a=\angle(\vec{r}_1,\vec{r}_2)+\angle(\vec{r}_2,\vec{r}_3)+\angle(\vec{r}_3,\vec{r}_1)$ between the Bloch vectors show inverse correlation with the entropy. 

\begin{table}
\begin{tabular}{|c|c|c|c|c|}
 \hline
$\angle S_a$ &  $\angle(\vec{r}_1, \vec{r}_{\rm  ss})$& $\angle(\vec{r}_2, \vec{r}_{\rm  ss})$ & $\angle(\vec{r}_3, \vec{r}_{\rm  ss})$ & $h$  \\ \hline
 235.489 & 115.323 & 2.42085 & 0.0313546&0.020 \\ 
221.528 & 109.238 & 3.5805 & 0.0390626 &0.023\\ 
 189.578 & 94.7316 & 5.87493 & 0.0575965 &0.026 \\
  26.9345 & 0.654795 & 12.8124 & 5.30314 &0.466\\
 14.2435 & 1.6635 & 5.45759 & 2.77901 &1.171\\
 13.0614 & 3.06505 & 1.4863 & 3.46569 &1.299\\ \hline
\end{tabular} 

\caption{\label{table1}
This table reports geometric characteristics of six distinct solutions for $\varepsilon=0.15$. It shows that the total angle (column 1)
for the ensemble shows the inverse correlation with the entropy $h$.
}
\end{table}

\subsection{Stability of three-state jumping}\label{Sec:Fluorostable3}

We now analyze the stability of the above 3-state jumping schemes.  
We do this using results from Secs.~\ref{Sec:Kjump}-\ref{Sec:stages} and numeric solutions for three-state jumping computed in Sec.~\ref{3stateJump}. As shown before, the stability of an individual trajectory is determined by the properties of the effective Hamiltonian $\hat{H}(\mu_i)$ from Eq.~(\ref{eq:Hwithmu}), where index $i$ ranges from 1 to 3 and $\mu_i$ are the settings for adaptive monitoring that generate three-state jumping for resonance fluorescence system. We extract values for $\mu_i$ using the expression derived in Ref.~\cite{KarWis11} for cyclic jumps that states that 
\begin{equation}
\hat{c} \ket{v^e_k} = -\mu_k\ket{v^e_k} + b_k\ket{v^e_{k+1}},
\label{eq:mu}
\end{equation}
where $k+1$ stands for $k+1\, \text{mod}\,3$ and $b_k$ is some constant. Once $\ket{v^e_k}$ and $\ket{v^e_{k+1}}$ are known, coefficients $\mu_k$ and $b_k$ are uniquely determined. Thus our approach for determining the stability of three-state jumping proceeds as follows. We convert the three-state jumping solutions in terms of Bloch vectors $\vec{r}_1$ $\vec{r}_2$ and $\vec{r}_3$  into state vectors $\ket{v^e_1}$, $\ket{v^e_2}$, and $\ket{v^e_3}$. Using Eq.~(\ref{eq:mu}), we compute $\mu_k$  and $\hat{H}(\mu_k)$ with $k=1,\ldots, 3$. At this point we can conclude that evolution stage from state $\ket{v^e_k}$ to state $\ket{v^e_{k+1}}$ is stable if state $\ket{v^e_k}$ has the smallest eigenvalue with respect to operator $\hat{H}(\mu_k)$. Otherwise, such stage is unstable. 
From the eigenvalues $\lambda^e_j$ and $\lambda^o_j$ one can determine mean-square stability using \erf{eq:KstateC}. 
 
\begin{figure}
 \includegraphics[width=.85\columnwidth]{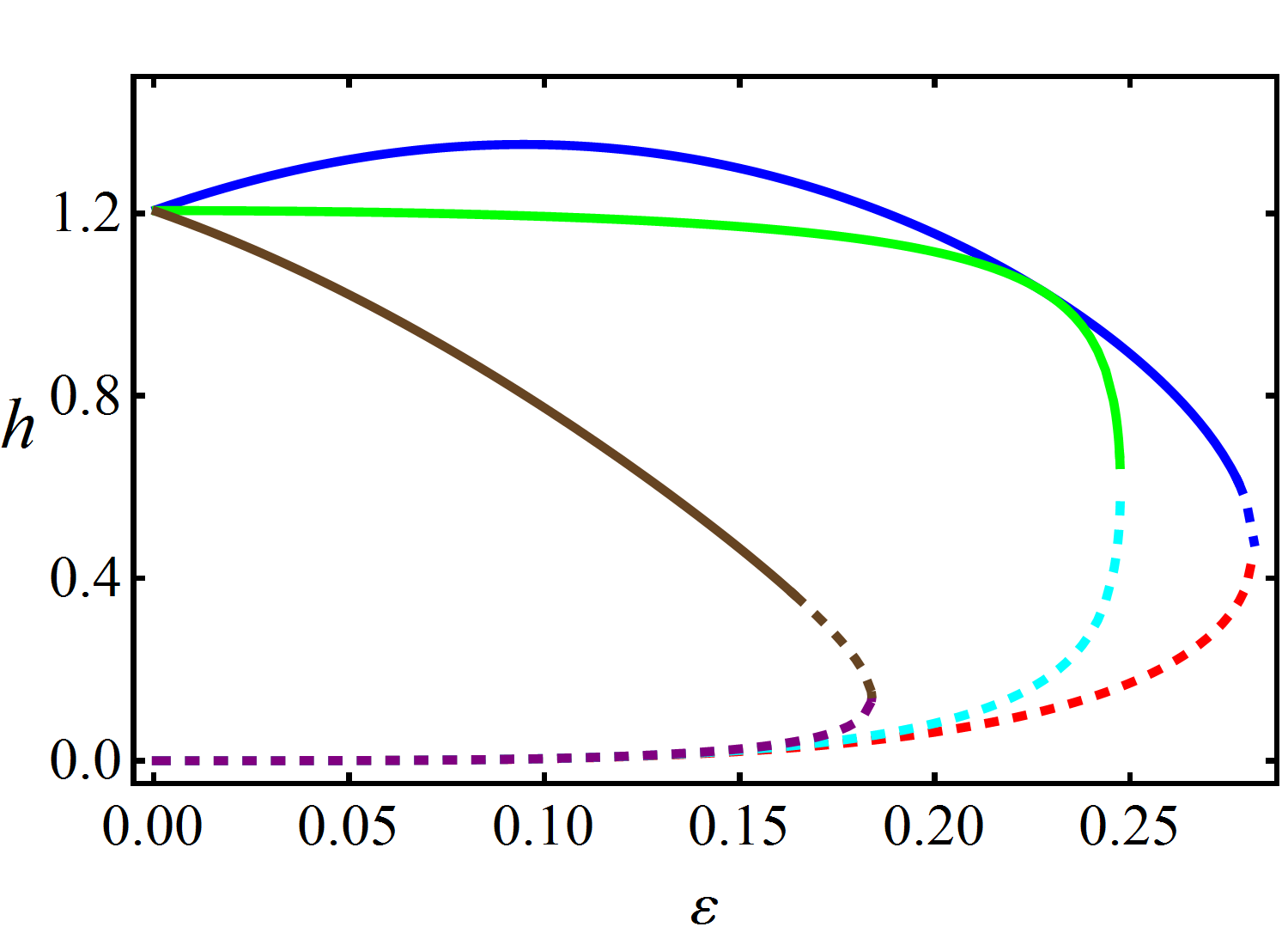} 
\vspace{-2mm}
\caption{\label{3sstab} (Color online.)  Ensemble Shannon entropy $h$ for the eight  different three-state jumping solutions (two of which come in degenerate pairs). The solid line indicate mean-square stability as well as piecewise deterministic stability. The  dashed line indicates that the solution has only mean-square stability, with one of its stages being unstable. }
\end{figure}

Results for this analysis are reported in Fig.~\ref{3sstab} and resemble the results for two-state jumping quite closely. Figure~\ref{3sstab} shows ensemble Shannon entropy $h$ for all possible three-state jumping solutions. Instability of individual trajectories for three-state jumping always arise in the same way as for two-state jumping: one of the three stages (corresponding to jumping from least probable state) becomes unstable. Solutions with small entropy (i.e., solutions with $h \to 0$ as $\varepsilon \to 0$) have mean-square stability, but not piecewise deterministic stability. Solutions with large entropy have both mean-square and piecewise deterministic stability, except in small region, where the entropy curves approach lower entropy solutions and develop one unstable stage in the evolution.

\section{Conclusion}
In this paper we have considered  a qubit undergoing evolution given by a Markovian master equation, subject to continuous monitoring that resolves every jump and allows the system to stay in a pure state. We studied special adaptive monitoring schemes that, in principle, allow an experimenter track the evolution of such system with finite-state machine as an apparatus. That is, the system jumps between only finitely many different states, the states in the associated {\em physically realizable ensemble}. The main contribution of this paper beyond Ref.~\cite{KarWis11} was analysis of the stability of such monitoring schemes. This is necessary to establish that the finite physically realizable ensembles introduced in Ref.~\cite{KarWis11} are not just mathematical constructions, but really are physically realizable. 

Because the evolution of the system is stochastic --- it undergoes jumps at random times --- we concentrated on the average properties of the system over many different realizations. Specifically, we derived conditions for the mean-square stability (that is, the long-time convergence of the fidelity of the system with the correct ensemble state). We showed that there is a positive parameter $C$, which is never more than unity, that guarantees mean-square stability as long it is not equal to unity. For the specific example of resonance fluorescence we considered 11 different finite-state monitorings (with two or three different states). In all cases $C$ was strictly less than unity (by a long way, in fact). Based on this, we conjecture that all finite physically realizable monitorings are mean-square stable. 
 However we also considered the long-time {\em rate} $R$ of convergence (which is exponential in time)  and found that some monitoring schemes converge much faster than others, in a way contrary to what is suggested by considering $C$ alone. 

Although all monitorings we considered were mean-square stable, there is a variety of behavior in the convergence of individual trajectories. Some monitorings give trajectories that are guaranteed to be piecewise deterministically stable (i.e. the between-jump evolution always increases the fidelity). However others do not --- in some deterministic (between jump) stages of the evolution the system moves {\em away} from the ensemble state that it ``should'' be in at that stage (and will be in, in the long time limit). Moreover, the stochastic jumps can contribute to both stability and instability of individual trajectories, even under piecewise deterministically stable monitorings. However in the long time limit, as the system approaches the ideal ensemble states in the mean-square sense, a fidelity decreases upon a jump become very unlikely. 

We illustrated these effects for the specific example of resonance fluorescence with an adaptively controlled weak local oscillator. 
Interestingly we found that those monitorings that are {\em not} piecewise deterministically stable tend to be those that are {\em most} 
stable in the mean-square sense  (by the measure of $C$ being  very small). Moreover, these are the monitorings that produce low entropy solutions, where 
the system spends most of its time in one state in the ensemble. Such monitorings would make it possible, in principle, to track a 
large number $N$ of qubits, using  much less than $N$ bits of memory (i.e. a finite state machine with far fewer than $2^N$ states). 

It is important to note that there are many open questions in this field of quantum state tracking with finite state machines.   First, given a physically realizable finite ensemble, does there exist an 
explicit construction for an adaptive monitoring scheme that 
realizes it? Second, is any such monitoring mean-square stable, as conjectured here? 
Third, is it true that any $D$-dimensional ergodic Markovian quantum system can be tracked by a $K$-state machine with some  finite $ K \geq (D-1)^2+1$, as conjectured in Ref.~\cite{KarWis11}. 
Fourth, is $K=(D-1)^2+1$ always sufficient? Fifth, can an example system be found that 
proves that $K=D$ is {\em not} sufficient in general ($D=3$ would be the minimum system size for such a search).
If the last can be proven, then there may be a relation to the recent result  that there are classical stochastic processes that can be generated using quantum systems of lower  entropy  than that required using only classical systems~\cite{GWRV}.
Finally, it seems likely that any given master equation would have additional structure that would enable 
one to use a smaller $K$ than that conjectured above, and this idea also remains to be explored.

\acknowledgments
This research was conducted by the Australian Research Council Centre of Excellence for Quantum Computation and Communication Technology (project number CE110001027).  We thank Mile Gu, Kurt Jacobs, and Peter Shor for discussions.

\appendix
\section{Useful Identities}
We now explain how to derive expressions for the convergence coefficient $C$ that determines how quickly the state  subject to adaptive monitoring approaches the states in the corresponding PR ensemble. We begin with the two-state jumping scenario. In this case, the expression for $C$ is given in Eq.~(\ref{eq:msF}). We evaluate it  through relating $Q^1_{11} Q^2_{11}$ and $Q^1_{22} Q^2_{22}$. We do this by comparing the  general expression  $\int d^n\vec\tau  |\beta_n|^2$ to $\int  d^n\vec\tau  |\alpha_n|^2$  for the special case when the initial state $\ket{\psi_0}$ is the jumping stat $\ket{v_1^e}$, i.e. $\alpha_0=1$ and $\beta_0=0$. Then with \erfs{tev1}{srel} and Eq.~(\ref{eq:unnormS}), we get 
\begin{align}
\alpha_{{ 2l}+1}&=\alpha_{{ 2l}}e^{-\lambda_1^e { \tau}_{{ 2l}+1}} Q^1_{11}\label{alpharel1}\\
\alpha_{{ 2l}}&=\alpha_{{ 2l}-1}e^{-\lambda_2^e { \tau}_{{2l}}}Q^2_{11}.\label{alpharel2}
\end{align}
Again by induction we deduce that 
\begin{align}
\alpha_{{ 2l}}&=(Q^1_{11} Q^2_{11})^l\prod_{j=1}^l e^{-\lambda_1^e { \tau}_{2j-1}}e^{-\lambda_2^e { \tau}_{2j}} \label{alpha2l} \\
\alpha_{{ 2l}+1}&= Q^1_{11}(Q^1_{11} Q^2_{11})^l\prod_{j=0}^l e^{-\lambda_2^e { \tau}_{2j}} e^{-\lambda_1^e { \tau}_{2j+1}}
\label{alpha2lp1}
\end{align}
Then 
\begin{equation}
\int d^n\vec\tau||\ket{\tilde{\psi}_{{ 2l}}}||^2 = \int d^n\vec\tau|\alpha_{{ 2l}}|^2  =\left(\frac{|Q^1_{11} Q^2_{11}|^2}{4 {\rm  Re}\lambda_1^e {\rm  Re}\lambda_2^e}\right)^l.
\label{eq:Qcond3}
\end{equation}
By construction $\int d^{2l}\vec\tau||\ket{\tilde{\psi}_{{ 2l}}}||^2 =1$. Therefore, Eq.~(\ref{eq:Qcond3}) proves that 
\begin{equation}
\frac{|Q^1_{11} Q^2_{11}|^2}{4 {\rm  Re}\lambda_1^e {\rm  Re}\lambda_2^e}=1.
\end{equation}
To calculate  the value of $C$, we relate $Q^1_{11} Q^2_{11}$ to $Q^1_{22} Q^2_{22}$. We recall the discussion after Eq.~(\ref{eq:jumpOp}), which showed that operator $\hat{s}_2\hat{s}_1$ is proportional to the identity operator. Using this fact and Eqs.~(\ref{srel}), we see that 
\begin{align}
\hat{s}_2\hat{s}_1 \ket{v_1^e}&=Q^1_{11} Q^2_{11} \ket{v_1^e} \label{eq:Q11} \\
\hat{s}_2\hat{s}_1 \ket{v_1^{{ o}}}&=Q^1_{22} Q^2_{22} \ket{v_1^{{ o}}}.\label{eq:Q22}
\end{align}
Since $\hat{s}_2\hat{s}_1$ is proportional to the identity, we can conclude $Q^1_{11} Q^2_{11}=Q^1_{22} Q^2_{22}$. 

This discussion can be easily generalized to the $K$-state jumping scenario.
 We again  calculate  the value of  $\int d^{nK}\vec\tau|\alpha_{n K}|^2$ in two ways  when $\alpha_0=1$ and $\beta_0=0$.  The first way is to note that, as before,  in this case  $\int d^{nK}\vec\tau|\alpha_{n K}|^2$  = 
 $\int d^{nK}\vec\tau||\ket{\tilde{\psi}_{nK}}||^2 = 1$.  The second way is to use the fact that in this case 
\begin{equation}
\alpha_{n K}= (Q^1_{11} \cdots Q^K_{11})^n\prod_{j=0}^{n-1} e^{-\lambda_1^e {\tau}_{jK+1}}\cdots e^{-\lambda_K^e { \tau}_{(j+1)K}},
\end{equation}
so that 
\begin{equation}
\int d^{nK}\vec\tau |\alpha_{n K}|^2 = \left(\frac{Q^1_{11}\cdots Q^K_{11}}{2^n {\rm  Re}\lambda_1^e \cdots {\rm  Re}\lambda_K^e}\right)^n.
\end{equation}
Thus ${Q^1_{11}\cdots Q^K_{11}} = {2^n {\rm Re}\lambda_1^e \cdots {\rm Re}\lambda_K^e}$. 
Next, we show  that  $Q^1_{11} \cdots Q^K_{11}=Q^1_{22}\cdots Q^K_{22}$. We  
prove this relationship for $Q^j_{kk}$ by considering full cycle jump operator $\hat{S}_K=\hat{s}_K\cdots\hat{s}_1$.
  
Assuming (as we can do without loss of generality) that the  jump operator $\hat{c}$ is traceless, we have $\hat{c}^2\propto \hat{I}$ so that  
\begin{equation}
 \hat{S}_K\ket{v_1^e}=[g_1(\mu_1,\ldots,\mu_K)+g_2((\mu_1,\ldots,\mu_K) \hat{c}]\ket{v_1^e}
\label{eq:gjumpC}
\end{equation}
The exact nature of functions $g_1$ and $g_2$ depend on the proportionality constant relating $\hat{c}^2$ to  the identity operator. Just as for Eq.~(\ref{eq:jumpOp}), a system can undergo jumping dynamics iff $g_2=0$.  This means that the full cycle jump operator $\hat{S}_K$ is proportional to the identity and we conclude that $Q^1_{11} \cdots Q^K_{11}=Q^1_{22}\cdots Q^K_{22}$, using the trick from \erfs{eq:Q11}{eq:Q22}.

\section{Groebner Basis}
\label{app:GroebnerBasis}
The task of finding jumping states for adaptive unravellings requires solving non-linear equations. We now introduce important concepts from computational algebraic geometry and review some algorithms used to solve multivariate polynomial systems. Our presentation will rely on analogies with linear algebra. 
From now on we regard a polynomial as the finite sum of terms, where each term is a product of a coefficient and monomial. 

Suppose we want to solve the system of non-linear equations,
\begin{equation} 
\{f_1(\vec{x}_n)=0, f_2(\vec{x}_n)=0, \ldots, f_s(\vec{x}_n)=0\},
\label{eq:nleql}
\end{equation}
where $\{f_1, f_2, \ldots, f_s\}$ are the polynomials with real rational coefficients 
and $\vec{x}_n=(x_1, x_2,\ldots, x_n)$ is the list of variables. 
Our goal is to find all the solution to the set of equations in~\erf{eq:nleql} and a concept of \emph{ideal} becomes useful. A collection of polynomials generates the ideal 
via $I=\left\langle f_1, f_2, \ldots, f_n\right\rangle=\left\{\sum_{i=1}^s h_i f_i\,:\,\, h_1, \dots, h_s\in \mathbb{C}[\vec{x}_n]\right\}$, where $s$ can be any finite index and $\mathbb{C}[\vec{x}_n]$ is a collection of all possible polynomials with complex coefficients with variables $\vec{x}_n$. Thus  ideals are similar to vector spaces, which are formed from all possible scalar combination of vectors. But instead of scalars, one uses all possible polynomial functions, $h_k$, defined on $\mathbb{C}[\vec{x}_n]$ to form an ideal. Ideals are important because the solution set to the newly created ideal and to the  original system are the same.

Different set of equations can have the same solution set and thus generate the same ideal. One of the main ideas in algebraic geometry is to pick a good set representing the ideal that has nice properties and yields easy way for identifying the solutions. Such  a set is called a \emph{Groebner basis}.  

Gaussian elimination is the algorithm used to solve system of linear equations. The extension to this algorithm used to solve a system of polynomial equations is known as Buchberger's algorithm and it is implemented in many packages for symbolic computations such as Mathematica, Maple, Sage. The set of equations obtained as a result of these algorithms is known as the Groebner basis. 

Before proceeding we explain how to check if a given set of equations produced by some software package is indeed a Groebner basis.  To do this, we first need to know the \emph{S-polynomial}. Given two polynomials $f$ and $g$, let $\mathbf{x}^\alpha$ be the least common multiple of leading terms of $f$ and $g$, denoted by $LT(f)$ and $LT(g)$. Then the $S$-polynomial is computed via $S(f,g)= f\xfrac{\mathbf{x}^\alpha}{LT(f)} -g \xfrac{\mathbf{x}^\alpha}{LT(g)}$. 
A set of polynomials $G=\{g_1, \ldots, g_n\}$ are a Groebner basis iff for all $i\neq j$, the remainder on division of $S(g_i, g_j)$ by $G$ is zero. An alternative explanation proceeds as follows. In general, one can always write  $S(g_i, g_j)$ as  $S(g_i, g_j)=\sum_{i=1}^n a_i g_i+r$, where $a_i$ are some polynomials and degree of the remainder polynomial $r$ is smaller than any polynomial in $G$.  Then $G$ is a Groebner basis iff the remainder polynomial $r$ is zero for all $i\neq j$. Division by a collection of polynomials is usually implemented in the software packages for computational algebraic geometry.

Now we explain how to use a Groebner basis to find \emph{all} solutions to the system of non-linear equations. Unlike linear algebra, where a vector space always has the same number of basis vectors, different Groebner bases can have  different number of elements and drastically different properties for the same ideal. The collection of  Groebner  bases arise from different orderings of monomials in the system of polynomials. 
Two particular orderings are relevant for the current discussion: \emph{lexicographic} order (Lex) and \emph{degree reverse lexicographic} order (DRL). Lexicographic order is
 an alphabetical order (write out monomial in full without any powers and order like the words in a 
dictionary from left to right). For DRL order, we first compare the total degree and then perform lexicographic 
ordering by reading the expressions from right to left. Both orderings allow different access to information 
about solutions to polynomial system of equations. The details on different orderings can be found in \cite{CLO}

Lex order is particularly useful because of the elimination theorem~\cite{CLO}, which states that the Groebner basis computed with respect to Lex order will yield a set of polynomials that can be solved by back substitution. This means that if the system of non-linear equations has finite number of solutions and the Lex  Groebner basis are given by $G=\{g_1, \ldots, g_t\}$, then $g_1=g_1(x_1)$ is a polynomial function of one variable only and we can determine all values for $x_1$ by solving $g_1(x_1)=0$. The next polynomial is most likely $g_2(x_1, x_2)$ is a function of two variables so that we can solve for $x_2$ given all possible values for $x_1$. This sequential substitution allows to solve for all variables $\vec{x}_n$. Some of the elements of $G$ are just constraints that eliminate possible solutions. Such procedure is used to solve a system of non-linear equations in Mathematica (command NSolve). 

This algorithm comes with some complications. For some systems, algorithm becomes unstable. This is so because the degree and size of coefficients in the Lex Groebner basis quickly become huge even for relatively small systems. And solving $g_1(x_1)=0$ involves rounding, which introduces errors that quickly propagate through back substitution and yield wrong results. 

The Groebner basis computed with respect to DRL ordering offers an alternative way to compute solutions to polynomial system of equations that does not involve back substitution. We now review machinery from algebraic geometry needed to compute solutions using DRL ordering of monomials. The first construct involves the space of all polynomials $\mathbb{C}[\vec{x}_n]$ and ideal $I$ defined above.  
Polynomials $\mathbb{C}[\vec{x}_n]$ can be classified into distinct categories (cosets) such that polynomials $f$ and $g$ belong to same coset iff $f-g$ have $0$ remainder with respect to ideal $I$ or equivalently polynomials $f$ and $g$ have the same remainder with respect to $I$. We denote the reminder of $f$ by $[f]$. Note that for checking this criteria it is sufficient just to consider remainder with respect to  the Groebner basis (same ordering as the one used for division algorithm) spanning $I$. 

We can consider the space of cosets, i.e., the space of the remainders with respect to division by $I$. This is a quotient ring, denoted by $\cal{A}=\mathbb{C}$~$[\vec{x}_n]$.  It also happens to be an algebra that has vector space structure and carries important information about solutions to the system of polynomials that spanned the ideal $I$. 

We now point the key steps in computing solutions to polynomials  $\{ f_1, f_2, \ldots, f_n\}$ and illustrate them using a simple example.
\begin{itemize}
 \item We compute the Groebner basis, $G=\{g_1, \ldots, g_t\}$,  with respect to DRL ordering for the ideal $I=\left\langle f_1, f_2, \ldots, f_n\right\rangle$. 
\item We identify leading term for every element of the Groebner basis $LT(g_i)$.
\item We create set $B$ from monomials that a not divisible by $LT(g_i)$, i.e., monomials in $B$ have degree smaller than the degree of $LT(g_i)$. This set is a basis for space of remainders, $\cal A$. 
\end{itemize}
For example, suppose we want to solve a system of equations
\begin{equation}
\left\{ \begin{array} {l}
 f_1=x^2-y^2+xy=0 \\
  f_2=x^2 y+y-1=0
       \end{array} \right.
\end{equation}
The Groebner basis (DRL ordering and $x>y$) for this system is
\begin{equation}
\left\{ \begin{array} {l}
 g_1=x^2+xy -y^2 \\
  g_2=y^3-x y^2+y-1\\
g_3=y^4+xy +2 y^2-x-2y
       \end{array} \right.
\end{equation}
Leading terms for these polynomials (with respect to chosen ordering) are $LT(I)=\{x^2, x y^2, y^4\}$ and the set $B=\{1, x, y, xy, y^2, y^3\}$ is the basis for Quotient ring. 

To extract the information about solutions from the quotient ring $\cal A$, we associate each polynomial $f$ in $\mathbb{C}[\vec{x}_n]$ with a linear map $m_f:\cal{A}\rightarrow \cal{A}$ whose action is given by $m_f(g)=[f g]$, where $g$ is some polynomial. Map $m_f$ can be represented as a matrix. To do this, we consider the action of $m_f$ on basis elements in the set $B$. And $m_f(b)$ for every $b\in B$ is an element of quotient ring $\cal A$ can be written as a column vector with respect to $B$.

For example discussed above we show how to compute $m_x$. The action of $m_x$ on $1$ is given by
\begin{equation}
 m_x(1)=[x]=(0,1,0,0,0,0)^T,
\end{equation}
i.e.,  $m_x$ is the coset $[x]$, which can be represented as a vector with respect to elements in basis set $B$. 
Map $m_x$ will take some basis elements outside of $B$. In this case, we compute the remainder with respect to the  Groebner basis and again compute a column vector with respect to $B$. For example, 
\begin{equation}
 m_x(x)=[x^2]=[g_1+y^2-x y]=[y^2-x  y]=(0,0,0, -1, 1,0)^T.
\end{equation}
Then the matrix form for $m_x$ is

\begin{equation}
m_x=\begin{pmatrix}
    0&0&0&1&-1&0\\
    1&0&0&0&0&1\\
    0&0&0&-1&1&1\\
    0&-1&1&0&0&-1\\
    0&1&0&0&0&-1\\
    0&0&0&1&1&0
   \end{pmatrix}
\end{equation}

We need to compute these matrices because they have very nice property: eigenvalues of $m_f$ are the values of polynomials  $f$ on solutions for original solutions. Thus eigenvalues of $m_x$  are all possible values a variable $x$ takes in the solutions set. Moreover, we don't need to compute every possible matrix $m_f$ because $m_{f g}=m_f m_g$. This is significant since computing $m_f$ relies on costly symbolic calculations, which are slow,  whereas matrix multiplication is fast. 

Thus eigenvalues for all different maps $m_{x_i}$ tells us all possible values each variable $x_i$ can take. But to learn the full solution (how to combine different $x_i$ to form $\vec{x}_n$) we also need information that is encoded in the eigenvector of $m_f$ for some $f$. Let $\bf{v}$ be such an eigenvector and we normalize this vector so that the first component is one (assuming that 1 is the first element in $B$). 
If variable $x_i$ is the $j$-th element in the set $B$, then $j$th component of $\bf v$ is the value of $x_i$ that yields one of the solutions. All values extracted in this way from one eigenvector come from one solution. 
However, this procedure does not work for every polynomial $f$. If $m_f$ has degenerate eigenvalues then corresponding eigenvectors cannot be used to extract the solution.  However, generic linear combination of the variables will yield a desired matrix. The details on this can be found in~\cite{Cox, CLO05}

For three state jumping, we computed the solutions using the Groebner basis with DRL order. The set of equations for cyclic three-state jumping has a lot of symmetry, e.g., one can map $\vec{r}_1\mapsto\vec{r}_2\mapsto\vec{r}_3\mapsto\vec{r}_1$ and $\kappa_{12}\mapsto\kappa_{23}\mapsto\kappa_{13}\mapsto\kappa_{12}$ 
and the system will remain unchanged. As the result, all matrices $m_{x_i}$ associated with unknowns in the system  have degeneracy with respect to real eigenvectors and cannot be used to extract the solution. Instead we used matrix $m_f$, where $f=r_{11}+\kappa_{12}$. Here $r_{11}$ is the first component of Bloch vector $\vec{r}_1$.  The reason for this choice is the following. Matrices associated with $\kappa_{12}$ and $\kappa_{23}$ and $\kappa_{31}$ are nice because they have least non-zero elements which speeds up the calculation (symbolic part). However, any combination of $\kappa_{12}$ and $\kappa_{23}$ and $\kappa_{31}$ or same coordinate of $r_1$ $r_2$, $r_3$ does not break the degeneracy due to symmetry above, but $r_{11}+\kappa_{12}$ does.

\end{document}